\documentclass[prd,10pt,aps,nofootinbib,twocolumn,floatfix,amssymb,amsmath,amsfonts]{revtex4-1}

\usepackage{epsfig}
\usepackage{slashed}
\usepackage{combelow} 
\usepackage{bm}

\usepackage[dvipsnames]{xcolor}
\definecolor{purple}{rgb}{0.8,0,0.6}

\newcommand{\beqn}{\begin{eqnarray}}
\newcommand{\eeqn}{\end{eqnarray}}
\newcommand{\beqs}{\begin{subequations}}
\newcommand{\eeqs}{\end{subequations}\\[-2mm]\noindent}
\newcommand{\eq}[1]{(\ref{#1})}

\newcommand{\cL}{{\cal L}}
\newcommand{\cS}{{S}}

\newcommand{\cD}{{\cal D}}

\newcommand{\PT}{{\mathcal{PT}}}

\newcommand{\Dirac}{\slashed D}
\newcommand{\dirac}{\slashed \partial}
\newcommand{\bs}{\boldsymbol}

\newcommand{\avr}[1]{{\left\langle #1 \right\rangle}}
\def\bbbone{{\mathchoice {\rm 1\mskip-4mu l} {\rm 1\mskip-4mu l} {\rm 1\mskip-4.5mu l} {\rm 1\mskip-5mu l}}}

\makeatletter
\DeclareRobustCommand{\cev}[1]{%
  {\mathpalette\do@cev{#1}}%
}
\newcommand{\do@cev}[2]{%
  \vbox{\offinterlineskip
    \sbox\z@{$\m@th#1 x$}%
    \ialign{##\cr
      \hidewidth\reflectbox{$\m@th#1\vec{}\mkern4mu$}\hidewidth\cr
      \noalign{\kern-\ht\z@}
      $\m@th#1#2$\cr
    }%
  }%
}
\makeatother

\begin{document}

\title{Anomalous dispersion, superluminality and instabilities in \\ two-flavour theories with local non-Hermitian mass mixing}

\author{Maxim N.~Chernodub}
\email{maxim.chernodub@univ-tours.fr}
\affiliation{Institut Denis Poisson, CNRS UMR 7013, Universit\'e de Tours, 37200 France}
\affiliation{Department of Physics, West University of Timi\cb{s}oara, Bd.~Vasile P\^arvan 4, Timi\cb{s}oara 300223, Romania}

\author{Peter Millington}
\email{peter.millington@manchester.ac.uk}
\affiliation{Department of Physics and Astronomy, University of Manchester, Manchester M13 9PL, United Kingdom}

%%%%%%%%%%%%%%%%%%%%

\begin{abstract}
Pseudo-Hermitian field theories possess a global continuous ``similarity'' symmetry, interconnecting the theories with the same physical particle content and an identical mass spectrum. In their regimes with real spectra, within this family of similarity transformations, there is a map from the non-Hermitian theory to its Hermitian similarity partner. We promote the similarity transformation to a local symmetry, which requires the introduction of a new vector similarity field as a connection in the similarity space of non-Hermitian theories. In the case of non-Hermitian two-flavour scalar or fermion mixing, and by virtue of a novel IR/UV mixing effect, the effect of inhomogeneous non-Hermiticity then reveals itself via anomalous dispersion, instabilities and superluminal group velocities at very high momenta, thus setting an upper bound on the particle momentum propagating through inhomogeneous backgrounds characterised by Lagrangians with non-Hermitian mass matrices. Such a non-Hermitian extension of the Standard Model of particle physics, encoded in a weak inhomogeneity of the non-Hermitian part of the fermion mass matrix, may nevertheless provide us with a low-energy particle spectrum consistent with experimentally observed properties.
\end{abstract}

\date{January 11, 2024}

\maketitle

%%%%%%%%%%%%%%%%%%%%

\section{Introduction}

Due to its wide applicability in experimental physics (see Refs.~\cite{Zyablovsky_2014, Christodoulides:2018arz, El-Ganainy:2018ksn, Ashida:2020dkc, 10.1093/nsr/nwy011, PRAVEENA2023171260} for reviews), pseudo-Hermitian quantum mechanics~\cite{Mostafazadeh:2001jk, Mostafazadeh:2001nr, Mostafazadeh:2002id}, wherein Hermiticity of the Hamiltonian is superseded by an antilinear symmetry such as parity--time-reversal ($\mathcal{PT}$)~\cite{Bender:1998ke, Bender:1998gh, Bender:2005tb} has inspired growing interest in viable quantum field theories (QFTs) with non-Hermitian Hamiltonians/Lagrangians. Examples include:\ non-Hermitian deformations of the Dirac Lagrangian with a parity-odd, anti-Hermitian mass term~\cite{Bender:2005hf,  Alexandre:2015kra, Alexandre:2017fpq, Alexandre:2017foi, Alexandre:2017erl} (see also Ref.~\cite{Jones-Smith:2009qeu}), theories of massive second-order fermions~\cite{LeClair:2007iy, Ferro-Hernandez:2023ymz}, scalar~\cite{Alexandre:2017foi, Alexandre:2017erl, Alexandre:2020gah, Alexandre:2020wki, Sablevice:2023odu} and fermionic~\cite{Alexandre:2015kra, Mishra:2018aej, Alexandre:2020wki} field theories with non-Hermitian mass mixing matrices, non-Hermitian Yukawa theories~\cite{Alexandre:2015kra, Alexandre:2020bet, Korchin:2021xxl, Mavromatos:2022heh, Croney:2023gwy}, scalar theories with complex~\cite{Bender:2004vn, Bender:2004sa, Bender:2012ea, Bender:2013qp, Shalaby:2017wux, Bender:2018pbv, Felski:2021evi, Ai:2022olh} or wrong-sign~\cite{Shalaby:2006fh, Shalaby:2009xda, Ai:2022csx} self-interactions,  theories exhibiting spontaneous symmetry breaking~\cite{Alexandre:2018uol, Mannheim:2018dur, Alexandre:2018xyy, Fring:2019hue, Alexandre:2019jdb, Fring:2019xgw, Fring:2020bvr} and topological defects~\cite{Fring:2020xpi, Fring:2021zci, Begun:2021wol, Begun:2022ufc, Correa:2022yjt}, holographic settings~\cite{Arean:2019pom, Morales-Tejera:2022hyq}, and non-Hermitian Dirac materials in the context of condensed matter physics~\cite{Juricic:2023szr}.

The attractiveness of non-Hermitian QFTs, whose spectra are nevertheless real and whose evolution is nevertheless unitary in regions of unbroken antilinear symmetry, stems from the unique phenomenology that they can exhibit. This unique phenomenology, in part, originates from the existence of so-called exceptional points, which mark the boundaries between regimes of broken and unbroken antilinear symmetry, and which cannot be reproduced by Hermitian theories.

In this work, we study one such unique phenomenology, which we previously identified in the scalar QFT composed of two complex scalar fields with a non-Hermitian mass mixing~\cite{Chernodub:2021waz}:\ that non-Hermitian QFTs with local Lagrangian parameters naturally lead to the emergence of a so-called ``similarity'' gauge field, momentum-dependent exceptional points, a new type of high-energy instability and a novel IR/UV mixing. Our aim here is to revisit these effects in scalar and fermionic theories with local non-Hermitian mass mixing matrices and to describe the associated phenomena of superluminal, negative or vanishing group velocities, and anomalous dispersion.
We note that time-dependent non-Hermitian quantum mechanical Hamiltonians have attracted significant attention~\cite{Fring:2020mhu, Fring:2021daw, Fring:2022qdp}. The phenomena described in this work are, however, unique to the case of relativistic field theories.

We focus on a model with two scalars and another with two fermions that have in common a similar non-Hermitian mass mixing, and highlight essential differences in the properties of these two models compared to one composed of a single Dirac fermion with a parity-odd, anti-Hermitian mass term. We consider local similarity transformations of the models and assume that the mass matrix has the same eigenvalues at every point in spacetime, thereby implying that the local ``low-energy'' measurements provide the same physical masses regardless of the time (spatial coordinate) when (where) this measurement is performed. Here, the term ``low-energy'' means that this statement is true for energies lower than the energy of typical ultra-high energy cosmic rays ($10^{18}$\,eV), thereby encompassing most observations to date. Similar to Ref.~\cite{Chernodub:2021waz}, we assume a spacetime dependence of the mass matrix, which, together with the fixed mass eigenvalues, corresponds to a spacetime-dependent rotation of the model in the ``similarity'' space.

While we find that the ground state of the single Dirac fermion becomes unstable, the low-energy modes, including the ground state, of the two-scalar and two-fermion models with non-Hermitian mass mixing are stable. The instability appears instead in the high-energy regime, implying potentially interesting consequences for the propagation of ultra-high-energy cosmic rays or for neutrino models with time- and space-varying mass matrices~\cite{Dvali:2021uvk}.

This article is structured as follows. We start our discussion in Sec.~\ref{sec:scalar} with a review and further update of the properties of the non-Hermitian model that describes a two-component complex-valued scalar field characterised by spatially inhomogeneous parameters. The local nature of the similarity map in this benchmark model allows us to introduce a novel kind of gauge field, dubbed the ``similarity gauge field'' in Ref.~\cite{Chernodub:2021waz}. In Sec.~\ref{sec:oneflavour}, we then follow the same strategy of introducing the similarity gauge field to a non-Hermitian fermionic model with a single flavour, finding that this single-flavour model gives us phenomenologically unacceptable results. In detail, we describe the theory of a single massive fermion with an anti-Hermitian, parity-odd mass term, involving the fifth gamma matrix. We promote its mass parameters to local functions and show that this straightforward generalisation leads to an unstable spectrum in the would-be $\mathcal{PT}$ unbroken regime.

In Sec.~\ref{sec:twoflavour}, we generalise the one-fermion model of Sec.~\ref{sec:oneflavour} to two fermion flavours, following the strategy successfully tested for the two-component scalar model in Sec.~\ref{sec:scalar}. We promote the mass parameters to local functions in Sec.~\ref{sec:local} and derive the dispersion relations for physical excitations in Sec.~\ref{sec:dispersion}, wherein we identify the high-momenta instabilities, anomalous dispersion and superluminal regimes, similar to the case of the doublet scalar model discussed at the beginning of the paper. The potential phenomenological relevance of our construction is described in Sec.~\ref{sec:pheno}. Section~\ref{sec:conc} outlines our conclusions, and additional technical details are provided in the Appendices.

%%%%%%%%%%%%%%%%%%%%

\section{Two-Flavour Scalar Model}
\label{sec:scalar}

In this section, we revisit the scalar model with local mass parameters, previously analysed in Ref.~\cite{Chernodub:2021waz}. Our aim is to show that, aside from the novel IR/UV mixing effect and resulting instabilities described in Ref.~\cite{Chernodub:2021waz}, certain modes also exhibit anomalous dispersion and superluminal propagation.

The Lagrangian of the model (originally introduced in Ref.~\cite{Alexandre:2017foi}) is
\begin{equation}\mathcal{L}_{\Phi}=\partial_{\mu}\tilde{\Phi}^{\dag}\partial^{\mu}\Phi-\tilde{\Phi}^{\dag}M^2\Phi\;,
\end{equation}
where $\Phi=(\phi_1,\phi_2)$ and
\begin{equation}
    M^2=\begin{pmatrix} m_1^2 & m_5^2 \\ -m_5^2 & m_2^2 \end{pmatrix}\neq M^{2\dag}
\end{equation}
is pseudo-Hermitian with
\begin{equation}
    \label{eq:scalar_P}
    P M^2 P=M^{2\dag}\;,\qquad P=P^{-1}=\begin{pmatrix} 1 & 0 \\ 0 & -1\end{pmatrix}\;.
\end{equation}
We take $m_1^2>m_2^2>0$ and $m_5^2>0$. The conjugate doublet $\tilde{\Phi}^{\dag}=(\tilde{\phi}_1^{\dag},\tilde{\phi}_2^{\dag})$ is defined via~\cite{Sablevice:2023odu}
\begin{equation}
    \tilde{\Phi}^{\dag}(x)=\eta^{-1}\Phi^{\dag}(x_{\eta})\eta\pi\;,
\end{equation}
where, for example, we can take $\eta=\mathcal{P}$ to be the parity operator and $\pi=P={\rm diag}(1,-1)$ to be the parity matrix, such that the complex field $\phi_1$ transforms as a scalar and the complex field $\phi_2$ as a pseudoscalar.\footnote{Since the field and its usual Hermitian conjugate evolve with the Hamiltonian $H$ and its Hermitian conjugate $H^{\dag}\neq H$, respectively, a Lagrangian formulated in terms of these variables would lead to inconsistent Euler--Lagrange equations~\cite{Alexandre:2017foi}. The discrepancy between the generator of time translations~\cite{Chernodub:2021waz} for the two fields would lead to further inconsistencies. Most notably, such a Lagrangian would not transform properly under the Poincar\'{e} group~\cite{Sablevice:2023odu}.} This choice matches Ref.~\cite{Alexandre:2020gah}, but it has the disadvantage that the momentum operator is no longer Hermitian and the operator $\eta$ is not sufficient for constructing a positive-definite inner product~\cite{Sablevice:2023odu}. Alternatively, we can take $\eta = \mathcal{P}\mathcal{A}$ and $\pi=P A$~\cite{Sablevice:2023odu}, where $\mathcal{A}$ is the additional discrete symmetry of the Lagrangian with
\begin{equation}
    A^{-1} M^2 A = M^2\;.
\end{equation}
Either way, since the equations of motion of this theory are linear, the exact choice does not impact the arguments presented in this work.

The squared mass eigenvalues are
\begin{equation}
    M_{\pm}^2=\frac{m_1^2+m_2^2}{2}\pm\sqrt{\left(\frac{m_1^2-m_2^2}{2}\right)^2-m_5^4}\;,
\end{equation}
and these are real when the parameter
\begin{equation}
    \zeta \equiv \frac{2m_5^2}{m_1^2-m_2^2}\leq 1\;.
\end{equation}
With these definitions, the matrix $A$ is given by
\begin{equation}
    \label{eq:scalar_A}
    A=\frac{1}{\sqrt{1-\zeta^2}}\begin{pmatrix} 1 & \zeta \\ -\zeta & -1 \end{pmatrix}\,,\quad PA=\frac{1}{\sqrt{1-\zeta^2}}\begin{pmatrix} 1 & \zeta \\ \zeta & 1 \end{pmatrix}\,.
\end{equation}
Note that $PAP=A^{\mathsf{T}}$ and $A^{-1}=A$.

The eigenvectors of the mass matrix are given by~\cite{Alexandre:2017foi}
\begin{equation}
\mathbf{e}_+=\begin{pmatrix} \cosh \xi_{\rm H} \\ -\sinh\xi_{\rm H}\end{pmatrix}\;,\qquad \mathbf{e}_-=\begin{pmatrix} \sinh \xi_{\rm H} \\ -\cosh\xi_{\rm H}\end{pmatrix}\;,
\end{equation}
where $\xi_{\rm H}=\frac{1}{2}{\rm arctanh}\,\zeta$. We can readily confirm that these eigenvectors are orthonormal with respect to the inner product
\beqn
\avr{{\bs a}, {\bs b}}_{{\mathcal{APT}}} = {\bs a}^{{\mathcal{APT}}} \cdot {\bs b} = {\bs a}^* \cdot P \cdot A \cdot {\bs b}\,,
\eeqn
 viz.
\beqn
\avr{{\mathbf{e}}_+, {\mathbf{e}}_+}_{{\mathcal{APT}}} & = & 
\avr{{\mathbf{e}}_-, {\mathbf{e}}_-}_{{\mathcal{APT}}} = 1\,, \\
\avr{{\mathbf{e}}_+, {\mathbf{e}}_-}_{{\mathcal{APT}}} & = &
\avr{{\mathbf{e}}_-, {\mathbf{e}}_+}_{{\mathcal{APT}}} = 0\,.
\eeqn

Since the squared mass matrix is non-Hermitian, it is diagonalised by a similarity transformation of the form
\begin{equation}
M_{\rm diag}^2=S_{\rm H}^{-1}M^2S_{\rm H}\;,
\end{equation}
where
\begin{equation}
\label{eq:scalar_similarity_matrix}
S_{\rm H}=\begin{pmatrix} \cosh \xi_{\rm H} & -\sinh \xi_{\rm H} \\ -\sinh\xi_{\rm H} & \cosh\xi_{\rm H}\end{pmatrix}\;.
\end{equation}
Note that the similarity matrix~\eq{eq:scalar_similarity_matrix} is related to the matrices \eq{eq:scalar_A} and \eq{eq:scalar_P} as follows:
\beqn
S_{\rm H} \cdot S_{\rm H} = A \cdot P\;.
\eeqn
In this way, the Hermitian Lagrangian
\begin{equation}
    \mathcal{L}_{\Phi,{\rm diag}}=\partial_{\mu}\Phi^{\dag}\partial^{\mu}\Phi-\Phi^{\dag}M^2_{\rm diag}\Phi
\end{equation}
corresponds to one of an infinite one-parameter family of isospectral Hamiltonians related via similarity transformations effected by the transformation
\begin{align}
\Phi&\to S\Phi\;,\\
\tilde{\Phi}^{\dag}&\to \tilde{\Phi}^{\dag}S^{-1}\;,
\end{align}
with
\begin{equation}
S=\begin{pmatrix} \cosh \xi & -\sinh \xi \\ -\sinh\xi & \cosh\xi\end{pmatrix}\;.
\end{equation}
That $\tilde{\Phi}^{\dag}$ is not the Hermitian conjugate of $\Phi$, except in the mass eigenbasis, is then manifest in the observation that $S^{-1}\neq S^{\dag}$.

The central idea of Ref.~\cite{Chernodub:2021waz} was to make the parameters $m_1^2=m_1^2(x)$, $m_2^2=m_2^2(x)$ and $m_5^2=m_5^2(x)$ spacetime dependent, and to generalise the global similarity transformation $S$ to a local similarity transformation with $\xi=\xi(x)$. It is then apparent that the kinetic term is not invariant under this local similarity transformation, but following the minimal coupling procedure, it is possible to restore invariance of the kinetic term by promoting the partial derivatives to covariant derivatives involving a ``similarity gauge field'' $C_{\mu}$, i.e.,
\begin{equation}
    \partial_{\mu}\to D_{\mu}=\partial_{\mu}-\mathcal{C}_{\mu}=\bbbone\, \partial_{\mu}+\sigma_1C_{\mu}\equiv \begin{pmatrix} \partial_{\mu} & C_{\mu} \\ C_{\mu} & \partial_{\mu} \end{pmatrix}\,.
\end{equation}
Under the similarity transformation, we have
\begin{equation}
\mathcal{C}_{\mu}\to S\mathcal{C}_{\mu}S^{-1}-S\partial_{\mu}S^{-1}\;.
\end{equation}
Recalling, however, that the tilde-conjugate field $\tilde{\Phi}$ transforms with $S^{-1}$, we also require the tilde-conjugate covariant derivative
\begin{equation}
    \tilde{D}_{\mu}=\partial_{\mu}+\mathcal{C}_{\mu}=\bbbone\, \partial_{\mu}-\sigma_1C_{\mu}\equiv \begin{pmatrix} \partial_{\mu} & -C_{\mu} \\ -C_{\mu} & \partial_{\mu} \end{pmatrix}\,.
\end{equation}
The similarity-invariant Lagrangian then takes the form
\begin{equation}
    \mathcal{L}_{\Phi,\mathcal{C}}=\tilde{D}_{\mu}\tilde{\Phi}^{\dag}D^{\mu}\Phi-\tilde{\Phi}^{\dag}M^2\Phi\;.
\end{equation}

We note that the diagonalisation of the coordinate-dependent mass matrix, effected by a coordinate-dependent similarity transformation that does not commute with the kinetic term, leads to the emergence of a non-vanishing similarity gauge field. Since the terms depending on the similarity gauge field are non-Hermitian, we therefore conclude that the non-Hermitian theory with local parameters cannot be mapped to a Hermitian theory, and it will consequently exhibit genuine non-Hermitian phenomena that a Hermitian theory cannot reproduce.

Following Ref.~\cite{Chernodub:2021waz}, we now consider the particular case in which the coordinate-dependent mass parameters $m_1^2$, $m_2^2$ and $m_5^2$ lead to coordinate-independent mass eigenvalues $M_{\pm}^2$. For a constant but non-vanishing similarity gauge field, the spectrum of the theory is then governed by
\begin{equation}
    \label{eq:scalar_characteristic}
    \left(p^2-C^2-m_1^2\right)\left(p^2-C^2-m_2^2\right)+4(C\cdot p)^2+m_5^4=0\;,
\end{equation}
where we have considered solutions to the Klein-Gordon equation of the form
\begin{equation}
    \Phi(x)=\varphi(p)e^{-ip\cdot x}\;.
\end{equation}
We adopt the standard conventions for the four-vector product $p \cdot x = p_\mu x^\mu = \omega t - {\bs p} \cdot {\bs x}$, where $p^\mu = (\omega, {\bs p})$ is the four-momentum and $x^\mu = (t,{\bs x})$ is the spacetime coordinate. We work with the mostly minus Minkowski signature. We have also introduced the following scalar Lorentz invariants:\ $p^2 \equiv p_\mu p^\mu = \omega^2 - {\bs p}^2$, $C \cdot p \equiv C_\mu p^\mu = C_0 \omega - {\bs C} \cdot {\bs p}$ and $C^2 \equiv C \cdot C \equiv C_\mu C^\mu \equiv C_0^2 - {\bs C}^2$.

A generic dispersion relation corresponds to a solution of Eq.~\eq{eq:scalar_characteristic} for a global, coordinate-independent vector $C^\mu$, which is given by a root of an algebraic equation of the 4th order. While such a solution is definitely possible to obtain in an analytical form, its rather complicated structure makes its further analytical analysis difficult. Therefore, we proceed below by considering two cases of strictly temporal or strictly spatial perturbations, where analytical solutions are provided by much simpler expressions. Moreover, these cases cover all possible variants for $C^\mu$, since a timelike or spacelike vector can be made, respectively, strictly temporal or strictly spatial with the help of Lorentz transformations.

For the case of a timelike similarity field $C^{\mu}=(C_0,\bm{0})$, the spectrum is given by
\begin{align}
    \omega_{\pm,\bm{p}}^2&=\bm{p}^2-C_0^2+\frac{m_1^2+m_2^2}{2}\nonumber\\&\phantom{=}\pm\sqrt{\left(\frac{m_1^2-m_2^2}{2}\right)^2-m_5^4-2C_0^2(m_1^2+m_2^2+2\bm{p}^2)}\;.
\end{align}
Noting that
\begin{align}
    m_1^2+m_2^2&=M_+^2+M_-^2\;,\\
    \left(\frac{m_1^2-m_2^2}{2}\right)^2-m_5^4&=\left(\frac{M_+^2-M_-^2}{2}\right)^2\;,
\end{align}
we can readily confirm that the energies $\omega_{\pm,\bm{p}}$ are coordinate independent. The modification to the eigenfrequencies leads to a corresponding modification of the positive definite inner product. This is described in Appendix~\ref{sec:innerprod}.

For sufficiently high momenta, the argument of the square root can become negative, such that the energy eigenvalues are real for momenta below some critical momentum $p_c$, and they come in complex-conjugate pairs for momenta above this critical momentum. In this way, the theory exhibits momentum-dependent exceptional points occurring at the critical momentum~\cite{Chernodub:2021waz}, given by
\begin{equation}
    p_c^2=\frac{\left(M_+^2-M_-^2\right)^2}{16C_0^2}-\frac{M_+^2+M_-^2}{2}\;.
\end{equation}

Proceeding similarly for a purely spacelike similarity field $C^{\mu}=(0,\bm{C})$, the spectrum is given by
\begin{align}
    \omega_{\pm,\bm{p}}^2&=\bm{p}^2-\bm{C}^2+\frac{m_1^2+m_2^2}{2}\nonumber\\&\phantom{=}\pm\sqrt{\left(\frac{m_1^2-m_2^2}{2}\right)^2-m_5^4-4\left(\bm{p}\cdot\bm{C}\right)^2}\;.
\end{align}
In this case, the instability arises for modes with a component $\bm{p}_\parallel$ parallel to $\bm{C}$ of magnitude greater than the critical momentum $p_{\parallel,c}$ given by
\begin{equation}
    p_{\parallel,c}^2=\frac{\left(M_+^2-M_-^2\right)^2}{16\bm{C}^2}\;.
\end{equation}

Having defined the critical momenta above, the timelike and spacelike cases can be expressed in the following convenient forms
\begin{equation}
    \omega_{\pm,\bm{p}}^2=\bm{p}^2+\overline{M}^2-\begin{cases} C_0^2\mp 2|C_0|\sqrt{p_c^2-\bm{p}^2}& \quad \text{timelike}\\ \bm{C}^2\mp 2|\bm{C}|\sqrt{p_{\parallel,c}^2-\bm{p}_{\parallel}^2}& \quad \text{spacelike}\end{cases}\;,
    \label{eq_omega_scalar_cases}
\end{equation}
where
\begin{equation}
\overline{M}^2=\frac{M_+^2+M_-^2}{2}
\label{eq_Mbar}
\end{equation}
is the average squared mass. Taking $C_0=C$ and $\bm{C}_i=C\delta_{i3}$ (along the $z$-direction), the group velocities take the form
\begin{equation}
    (\bm{v}_{\pm,\bm{p}})_i=\frac{1}{\omega_{\pm,\bm{p}}}\left[\bm{p}_i\mp C\begin{cases} \frac{\bm{p}_i}{\sqrt{p_c^2-\bm{p}^2}}&\quad \text{timelike}\\\frac{p_3\delta_{i3}}{\sqrt{p_c^2-{p}_3^2}}&\quad \text{spacelike}
    \end{cases}\right]\;.
\end{equation}
For modes propagating in the $z$ direction, $\bm{p} = (0,0,p)$, and the group velocities take the same form for the timelike and spacelike cases, i.e.,
\begin{equation}
    v_{\pm,p}=\frac{p}{\omega_{\pm,p}}\left[1\mp \frac{C}{\sqrt{p_c^2-p^2}}\right]\;.
\end{equation}

There exists another set of critical momenta $p_{\rm stop}$ and $p_{\parallel,{\rm stop}}$ at which the group velocity of the $\omega_+$ mode vanishes. These critical momenta are given by
\begin{equation}
p_{(\parallel,){\rm stop}}^2=p_{(\parallel,)c}^2-C^2\;.
\end{equation}
Moreover, for modes with the momenta in the intermediate range, $p_{(\parallel,){\rm stop}}^2<\bm{p}^2_{(\parallel)}<p_{(\parallel,)c}^2$, the group velocity of the $\omega_+$ mode is negative.

In standard optical materials, the index of refraction $n$ increases with increasing frequency $\omega$ (so that $\partial n/\partial \omega > 0$). This corresponds to ``normal'' dispersion. The latter is responsible, e.g., for the usual ordering of the colours in a rainbow, from red at the top to violet at the bottom. In the case of anomalous refractive index, $\partial n/\partial \omega < 0$, the order of colours would get reversed, from violet at the top to red at the bottom. 

Since the ``$\pm$''-modes possess different dispersions, they will have different refractive indices $n_{\pm,p}$ determined with respect to the corresponding phase velocities $u_{\pm,p} = \omega_{\pm,p}/p$ as follows:
\begin{align}
	n_{\pm,p} = \frac{1}{u_{\pm,p}} \equiv \frac{p}{\omega_{\pm,p}}\,.
	\label{eq_n_pm}
\end{align}
(We set the velocity of light to unity, $c = 1$.) The condition for the anomalous refractive index then reads as
\begin{align}
	\frac{\partial n_{\pm,p}}{\partial p} \equiv \frac{\partial }{\partial p} \frac{p}{\omega_{\pm,p}} < 0\,, 
	\quad
	\text{[anomalous dispersion]}\,.
	\label{eq_anomalous}
\end{align}
The dispersion of the ``$+$'' mode is always normal. However, the dispersion of the ``$-$'' mode becomes anomalous at a scale $p_a$ below both $p_c$ and $p_{\rm stop}$. For completeness, the scale $p_a$ is given by
\begin{align}
    p_a^2&=2p_c^2+\overline{M}^2-\frac{C^2}{2}-\frac{\overline{M}^4}{2C^2}\nonumber\\&\phantom{=}+\frac{\overline{M}^2-C^2}{2C^2}\sqrt{\left(\overline{M}^2-C^2\right)^2-4C^2p_c^2}\;,
\end{align}
where we have assumed $C^2 < M^2$, needed to preserve the real-valuedness of the dispersion for ``$-$'' mode at the vanishing momentum, ${\bs p} = 0$ as follows from Eq.~\eq{eq_omega_scalar_cases}.

The classical relation between the dispersion $n$ of a medium and the group and phase velocities
\begin{align}
	v_{\mathrm{gr}} = \frac{1}{n + \omega \frac{\partial n}{\partial \omega}} \,, \qquad\ n = \frac{1}{v_{\mathrm{ph}}} = \frac{p}{\omega}\,,
	\label{eq_v_gr_n}
\end{align}
implies that the group velocity may exceed the speed of light if the anomalous dispersion becomes too extreme, with $n + \omega \partial n/\partial \omega < 1$. Indeed, both modes $\omega_{\pm}$ exhibit superluminal propagation for momenta above the corresponding scales
\begin{equation}
    p^2_{\rm SL,\pm}=\frac{p_c^2}{\overline{M}^4}\left[\overline{M}^4\pm 2 C^2 p_c \sqrt{p_c^2+\overline{M}^2}-C^2(2p_c^2+\overline{M}^2)\right]\;,
\end{equation}
where $p_{\rm SL}<p_c$. Note that the superluminal region for the mode $\omega_+$ occurs when its group velocity is negative, such that $p_{{\rm stop}}<p_{{\rm SL},+}<p_{c}$.

In Sec.~\ref{sec:twoflavour}, we will see that a fermionic model with non-Hermitian two-flavour mixing exhibits analogous behaviours, wherein we provide a more comprehensive exposition of these uniquely non-Hermitian phenomena. A complete description of the analogous phase diagram and the critical points is given  for this phenomenologically interesting model in Sec.~\ref{sec:dispersion}.

%%%%%%%%%%%%%%%%%%%%

\section{One-Flavour Fermionic Model}
\label{sec:oneflavour}

Before considering the fermionic analogue of the two-flavour scalar model of the previous section, we focus on a simple non-Hermitian extension of the Dirac Lagrangian~\cite{Bender:2005hf,Alexandre:2015kra,Alexandre:2017fpq,Alexandre:2017foi,Alexandre:2017erl}, which has the following form:
\beqn
\label{eq_cL}
\cL_{\psi} = \bar\psi \left(i \dirac - m - m_5 \gamma^5 \right) \psi\;,
\label{eq_L:old}
\eeqn
where $m$ corresponds to the Hermitian mass term for the fermion field $\psi$, and $m_5$ provides us with the anti-Hermitian mass term. Note that the pseudo-Hermiticity of the Lagrangian necessitates a redefinition of the dual field $\bar{\psi}$, which is therefore not the usual Dirac-conjugate $\bar{\psi}\neq \psi^{\dag}\gamma^0$ (see Refs.~\cite{Alexandre:2022uns, Sablevice:2023odu}, cf.~Refs.~\cite{LeClair:2007iy, Ferro-Hernandez:2023ymz} in the case of second-order fermions).\footnote{As noted for the scalar case, we remark that the precise definition of $\bar{\Psi}$ has no impact on the dispersion relations relevant to this work, since the fermionic theory is linear.}

The Lagrangian~\eq{eq_cL} gives the following classical equations of motion:
\beqn
\left(i \dirac - m - m_5 \gamma^5 \right) \psi & = & 0\,, \\
\bar{\psi}\left(i \overleftarrow{\dirac} + m + m_5 \gamma^5 \right) & = & 0\,.
\eeqn
By inspection, we see that the dual spinor is defined in terms of the spinor $\tilde{\psi}^{\dag}$ and not $\psi^{\dag}$, where these differ by $m_5\to-m_5$ (see Ref.~\cite{Alexandre:2022uns}).

The positive-frequency solutions of the Dirac equation
\beqn
\psi(x) = u(p) e^{- i p\cdot x}
\eeqn
are expressed via the spinor $u(p)$,\footnote{The explicit form of the four-spinor can be found in Ref.~\cite{Alexandre:2022uns}.} which satisfies
\beqn
\left({\slashed p} - m - m_5 \gamma^5 \right) u(p) = 0\;.
\label{eq_Dirac:p}
\eeqn

The self-consistency of Eq.~\eq{eq_Dirac:p} requires $p^2 = M^2$ and determines the energy spectrum ($p_0 \equiv E_{\bs p}$) via
\beqn
\omega_{\bs p}^2 = {\bs p}^2 + M^2\;,
\eeqn
in which the squared mass of the fermionic excitation is
\beqn
M^2 = m^2 - m_5^2\;.
\label{eq_M:0}
\eeqn
We always take $|m| \geqslant |m_5|$ to ensure that the mass $M$ is a real quantity. This range of parameters corresponds to the ``$\PT$-symmetric'' phase, in which the theory is stable. If the Hermitian mass $|m|$ is smaller than the non-Hermitian mass $|m_5|$, then the system resides in the ``$\PT$-broken'' phase, which is characterised by paired complex branches of fermionic energies that make the vacuum unstable.

This theory, in fact, belongs to a one-parameter family of similar theories. This can be expressed conveniently by writing
\beqn
m + m_5 \gamma^5 = M e^{2 \gamma^5 \theta}\;,
\eeqn
with 
\beqn
m = M \cosh 2 \theta 
\qquad \text{and}\qquad 
m_5 = M \sinh 2 \theta\;,
\eeqn
where $\theta$ is a real parameter, and the mass of the fermionic excitations $M$ is given in Eq.~\eq{eq_M:0}. This family is connected by a corresponding non-unitary and non-compact similarity transformation that takes the form
\beqn
\psi & \to e^{\omega_5 \gamma^5} \psi\qquad \text{and}
\qquad
{\bar \psi} \to {\bar \psi}\,e^{\omega_5 \gamma^5}\;.
\label{eq_U1:S}
\eeqn

Applying the similarity transformation~\eq{eq_U1:S} with the constant parameter $\omega_5 = - \theta$ to the original non-Hermitian Lagrangian~\eq{eq_cL}, we obtain the corresponding Hermitian Hamiltonian
\beqn
\cL_{\psi, {\rm H}} = \bar\psi \left(i \dirac - M \right) \psi.
\label{eq_cL:H}\;
\eeqn
Notice that the kinetic term of the Dirac Lagrangian is invariant under the similarity transformation~\eq{eq_U1:S}.

We now assume that both the Hermitian and non-Hermitian masses in the original Lagrangian~\eq{eq_cL} are functions of the spacetime coordinate $x^{\mu}$, i.e., $m = m(x)$ and $m_5 = m_5(x)$, such that
\beqn
m(x) + m_5(x) \gamma^5 = M e^{2 \gamma^5 \theta(x)}.
\eeqn
In this way, the mass of the fermionic excitation~\eq{eq_M:0} remains a coordinate-independent quantity. Still, the original non-Hermitian Lagrangian~\eq{eq_cL} can no longer be mapped to its Hermitian analogue~\eq{eq_cL:H}, as we will see below.

In order to incorporate the coordinate dependence of the mass parameters $m$ and $m_5$, we need to promote the global similarity transformation~\eq{eq_U1:S} to a local transformation. To this end, we introduce the new axial vector similarity gauge field $C_\mu$ and promote the Lagrangian to
\beqn
\cL_{\psi,C} = \bar\psi \left(i \dirac + i \slashed{C} \gamma^5  - m - m_5 \gamma^5 \right) \psi\;.
\label{eq_cL:C}
\eeqn
Under the local version of the similarity transformation in Eq.~\eq{eq_U1:S}, the similarity gauge field transforms as
\beqn
C_\mu & \to C_\mu - \partial_\mu \omega_5\;.
\label{eq_U1:C:S}
\eeqn
The parameter $\omega_5$ originates from a non-compact gauge group. The similarity gauge field $C_\mu$ is, therefore, a non-compact gauge field, which, e.g., will not contain Abelian monopole-like singularities.

We now turn to the case of a constant similarity gauge field. For a purely spacelike field $C^\mu = (0, {\bs C})$, the spectrum is
\beqn
\omega^2_{\pm,\bs{p}} & = & m^2 - m_5^2 + {\bs p}^2 - {\bs C}^2 \nonumber \\
& & \pm 2 i\sqrt{(m^2 - m^2_5) {\bs C}^2 + ({\bs p} \cdot {\bs C})^2}\;.
\label{eq_omega:spatial}
\eeqn
While this spectrum functionally resembles the phenomenologically interesting energy dispersion of the scalar doublet model~\eqref{eq_omega_scalar_cases}, it possesses both unstable and dissipative modes at any momentum $\bs p$. This statement applies even for the $\mathcal{PT}$-broken regime with $m^2 - m^2_5 > 0$. The complex-valuedness of the energy spectrum renders this model of less interest for particle phenomenology.

\begin{figure}[!ht]
\begin{center}
\includegraphics[width=0.4\textwidth,clip=true]{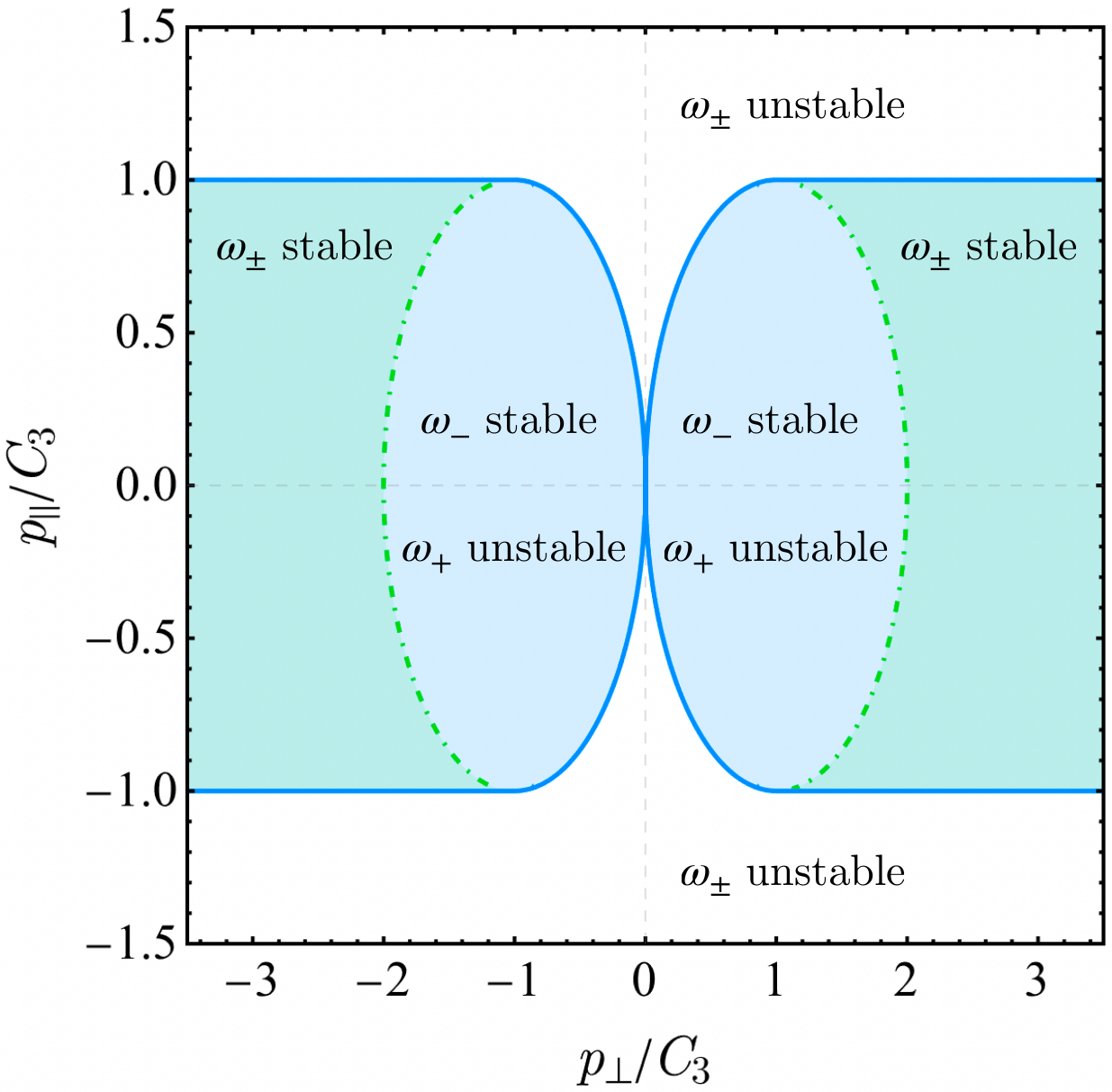}
\end{center}
\caption{An example of the stability regions of the one-fermion spectrum~\eq{eq_omega:spatial} in the plane of longitudinal $p_\| \equiv p_3$ and transverse \smash{$p_\perp = \pm \sqrt{p_1^2 + p_2^2}$} components of momenta for  $m_5^2 - m^2 = C_3^2 > 0$. The stability island of the $\omega_+$ mode (shown in green) lies within a more extended region of stability for the $\omega_-$ mode (shown in blue). In the uncoloured region, both modes are unstable.}
\label{fig_stabilisation}
\end{figure}

Unexpectedly, the spectrum~\eq{eq_omega:spatial} nevertheless contains extended ``islands of stability'' in momentum space in the would-be $\mathcal{PT}$-broken regime for which $m^2 - m_5^2 <0$. In this case, the spectrum always has an unstable low-momentum region as both $\pm$ branches of the squared energy dispersion~\eq{eq_omega:spatial} are negative at vanishing momentum, i.e., $\omega^2_{\pm}(\bs{p} = 0) < 0$. However, higher-momentum modes become stable at higher momenta. To illustrate these properties, we take a $\mathcal{PT}$-broken set of masses, i.e.,  $m_5^2 - m^2 = C_3^2 > 0$, with the vector similarity field pointing along the $z$ axis, such that ${\bs C} = (0,0,C_3)$. 
The stability region is given by the overlap of the two shaded regions of Fig.~\ref{fig_stabilisation}, which correspond to the stable domains for the $\omega_+$ and $\omega_-$ modes, respectively.
In this way, we find that this model exhibits momentum-dependent exceptional points, occurring along the boundaries of these regions of stability, as were found for the case of the non-Hermitian scalar field theory with local mass parameters~\cite{Chernodub:2021waz}, reviewed in Sec.~\ref{sec:scalar}.

For a purely timelike field $C^\mu = (C_0, {\bs 0})$, the spectrum is instead
\beqn
\omega^2_{\pm,\bs{p}} = m^2 - m_5^2 + \left(|{\bs p}| \pm i C_0 \right)^2\;.
\eeqn
For mass parameters in the physical $\mathcal{PT}$-unbroken regime ($m^2 > m_5^2$), these spectra share the same property:\ they include four complex eigenmodes, which are pairwise related to each other by complex conjugation. In both cases, the energy branches at nonzero momentum, ${\bs p}^2 >0$, always include both dissipative and unstable modes, which is an unacceptable feature for a phenomenological appropriate model.

Summarising this section, we stress that the IR instabilities of the one-flavour fermionic model equipped with the similarity gauge field seriously limit its phenomenological viability. In the next section, however, we will show that the two-flavour generalisation of the single-fermion model does not exhibit the same problems at low momenta and, therefore, can be useful in the non-Hermitian extensions of the Standard Model of particle physics.

%%%%%%%%%%%%%%%%%%%%

\section{Two-flavour fermionic model}
\label{sec:twoflavour}

In this section, we consider the following two-flavour generalisation of the non-Hermitian Dirac Lagrangian:
\beqn
\cL_{\Psi} = \sum_{a=1}^2 {\bar \psi}_a \left(i \dirac - m_a \right) \psi_a - m_5\left({\bar \psi}_1 \gamma^5\psi_2 + {\bar \psi}_2 \gamma^5\psi_1\right)\;.\nonumber\\
\label{eq_L:explicit}
\eeqn
We take $m_1>m_2>0$ and $m_5>0$ without loss of generality. The model~\eq{eq_L} mimics closely its bosonic counterpart with two species~\cite{Alexandre:2017foi}; in fact, the two systems can be considered supersymmetric partners~\cite{Alexandre:2020wki}.

\begin{figure}[!ht]
\begin{center}
\includegraphics[width=0.4\textwidth,clip=true]{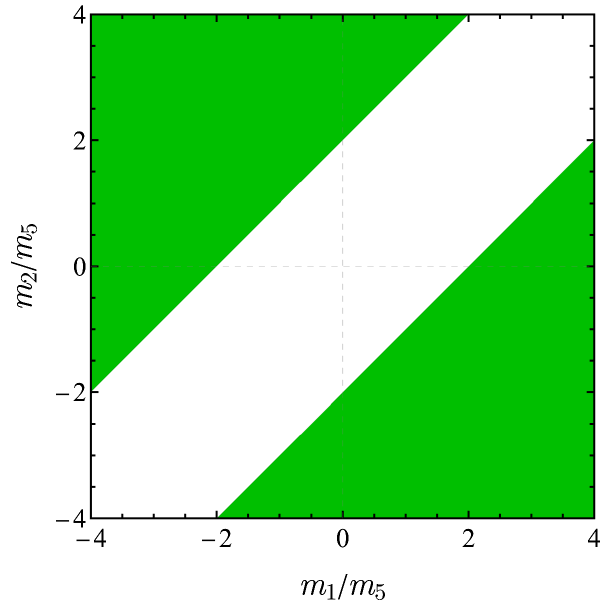}
\end{center}
\caption{The $\PT$-unbroken regions (shown in the green colour) for the model~\eq{eq_L} with the mass matrix~\eq{eq_mass_matrix} at $m_5 \neq 0$.}
\label{fig_PT_unbroken_region}
\end{figure}

If we take $\psi_1$ to be right-chiral and $\psi_2$ to be left-chiral, such that $\gamma^5\psi_1=+\psi_1$ and $\gamma^5\psi_2=-\psi_2$, the Lagrangian can be written in the more convenient form
\beqn
\cL_{\Psi} = \bar \Psi \left(i \dirac - {\hat M} \right) \Psi\;,
\label{eq_L}
\eeqn
where $\Psi = (\psi_1,\psi_2)^T$ is the doublet of fermionic fields, and the mass matrix takes the form 
\beqn
{\hat M} = 
\left(
\begin{array}{rr}
m_1 & \phantom{-} m_5 \\
- m_5 & m_2 
\end{array}
\right)\;.
\label{eq_mass_matrix}
\eeqn
Its eigenvalues are
\beqn
M_\pm = \frac{1}{2} \left( m_1 + m_2 \pm \sqrt{(m_1 - m_2)^2 - 4 m_5^2} \right)\;,
\label{eq_mass:eigenvalues}
\eeqn
and the energy dispersion relations take the standard relativistic form
\beqn
\omega_{\pm,\bs{p}}^2 = {\bs p}^2 + M_\pm^2\;,
\label{eq_omega_pm}
\eeqn
implying that Eq.~\eq{eq_mass:eigenvalues} represents the physical masses of the excitations in the system. The $\PT$-unbroken domain of the model~\eq{eq_L} occurs when the mass parameters satisfy
\beqn   
4 m_5^2 < (m_1 - m_2)^2\,,
\label{eq_PT_region}
\eeqn
which is illustrated in Fig.~\ref{fig_PT_unbroken_region}.\footnote{ Notice that a negative value of a fermionic (as well as bosonic) mass does not have any physical consequences because the fermions (bosons) with the masses $m$ and $-m$ have the same energy dispersions~\eq{eq_omega_pm}.}

Similarly to the two-scalar case, it is convenient to introduce the non-Hermiticity parameter
\beqn
    \zeta \equiv \frac{2 m_5}{m_1 - m_2}\;,
\label{eq_eta}
\eeqn
which determines the deviation of the model from the Hermitian point. Exactly at a vanishing value $\zeta=0$, the two fermionic flavours decouple, and the model becomes Hermitian. For $0 < |\zeta| < 1$, the model is non-Hermitian but resides in the domain of unbroken $\PT$ symmetry, where the mass eigenvalues are real. For $|\zeta| >1$, the mass eigenvalues~\eq{eq_mass:eigenvalues} are complex, and the $\PT$-symmetry is broken. At the points $\zeta = \pm 1$, the squared mass eigenvalues merge, and the mass matrix becomes defective. At this exceptional point, which occurs at the boundary between the regimes of broken and unbroken $\PT$ symmetry, the fermions become degenerate, and the model acquires a global $U(2)$ symmetry.

%%%%%%%%%%%%%%%%%%%%

\subsection{Global similarity transformation}
\label{sec:globalcase}

In the $\PT$-unbroken regime, the two-fermion model~\eq{eq_L} with spacetime-independent mass matrix can be mapped, via a global similarity transformation, to a Hermitian model. In order to demonstrate this property, it is convenient to consider the following parametrisation of the mass matrix~\eq{eq_mass_matrix}:
\beqs
\beqn
m_1 & = & M + m \cosh 2 \varkappa\,, \\
m_2 & = & M - m \cosh 2 \varkappa\,, \\
m_5 & = & m \sinh 2 \varkappa\,,
\label{eq_m5_param}
\eeqn
\label{eq_m:parameterization:const}
\eeqs
which reduces Eq.~\eq{eq_mass_matrix} to
\beqn
{\hat M} = M \left(
\begin{array}{rr}
1 & 0 \\
0 & 1
\end{array}
\right)
+ m \left(
\begin{array}{rr}
\cosh 2 \varkappa & \phantom{-} \sinh 2 \varkappa \\
- \sinh 2 \varkappa & \cosh 2 \varkappa 
\end{array}
\right).
\label{eq_m_hat}
\eeqn
In this parametrisation, the physical eigenmasses~\eq{eq_mass:eigenvalues} take the following simple form:
\beqn
M_\pm = M \pm m\,,
\label{eq_mass:eigenvalues:2}
\eeqn
where, according to Eq.~\eq{eq_mass:eigenvalues}, 
\begin{align}
    M = \frac{m_1 + m_2}{2}
    \quad \text{and}\quad 
    m = \frac{1}{2} \sqrt{(m_1 - m_2)^2 - 4 m_5^2}\;.
    \label{eq_mass:eigenvalues:3}
\end{align}
It is important to notice that the physical spectrum~\eq{eq_mass:eigenvalues:2} does not depend on the parameter $\varkappa$ that labels the similarity degeneracy of the non-Hermitian model. In the $\PT$-unbroken regime, the mass parameter $m$ is a real-valued quantity, whereas, in the broken phase, $m$ acquires an imaginary contribution. Notice that the standard non-Hermiticity parameter $\zeta$ defined in Eq.~\eq{eq_eta} is related to the parameter $\varkappa$ as $\zeta = \tanh 2 \varkappa$. The exceptional points $\zeta = \pm 1$ correspond to the asymptotic limits $\varkappa \to \pm\infty$, respectively.

Notice that the model~\eq{eq_L} becomes Hermitian for purely imaginary $\varkappa = \pm i |\varkappa|$ (with $M$ and $m$ real), since the off-diagonal terms become
$i \left({\bar \psi}_1 \psi_2 - {\bar \psi}_2 \psi_1\right)$. Consequently, the mass eigenvalues
\beqn
M^{({\rm Herm})}_\pm = \frac{1}{2} \left(m_1 + m_2 \pm \sqrt{(m_1 - m_2)^2 + 4 |m_5|^2} \right)\nonumber\\
\label{eq_mass:eigenvalues_H}
\eeqn
are real-valued for all values of the mass parameters, in contrast to their non-Hermitian counterparts~\eq{eq_mass:eigenvalues}.

We now turn our attention to the global similarity transformation
\beqn
\begin{array}{l}
{\mathrm{Global}} \\ 
{\mathrm{similarity}} \\
{\mathrm{transform}}
\end{array}
\!:\quad 
\left\{
\begin{array}{lcl}
\Psi & \to & \cS \Psi \\ 
\bar \Psi & \to & \bar \Psi \cS^{-1}
\end{array}\right.\;,
\label{eq_similarity:transformation}
\eeqn
which maps  the non-Hermitian Lagrangian~\eq{eq_L} into a Hermitian one of the form
\beqn
\cL_{\Psi,{\rm diag}} = \bar \Psi \left(i \dirac - {\hat M}_{\rm diag} \right) \Psi\;,
\label{eq_L:Hermitian}
\eeqn
containing the diagonalised mass matrix
\beqn
{\hat M}_{\rm diag} \equiv {\mathrm{diag}}(M_+,M_-) = \cS^{-1} {\hat M} \cS\;.
\label{eq_mass_matrix_diag}
\eeqn
The similarity transformation is represented by a $SU(1,1)$ matrix $\cS$, which operates in the isospace diagonal in the spinor space. It takes the explicit form
\beqn
\cS \equiv \cS(\varkappa) = e^{- \varkappa \sigma_1} \equiv \begin{pmatrix} 
\phantom{-} \cosh \varkappa & - \sinh \varkappa \\[1mm]
- \sinh \varkappa & \phantom{-} \cosh \varkappa 
\end{pmatrix}\,.
\label{eq_similarity:matrix}
\eeqn

Notice that for a real-valued parameter $\varkappa$ (i.e., in the $\PT$-unbroken regime), $S$ is Hermitian but not unitary, i.e., $\cS^\dagger \equiv \cS $ and $\cS^\dagger \neq \cS^{-1}$, with
\beqn
\cS^{-1} \equiv \cS(-\varkappa) = e^{\varkappa \sigma_1} \equiv \begin{pmatrix} 
\cosh \varkappa & \sinh \varkappa \\[1mm]
\sinh \varkappa & \cosh \varkappa 
\end{pmatrix}\,.
\label{eq_similarity:matrix_inv}
\eeqn
The multiplication rule 
\beqn
\cS(\varkappa_1)\cdot\cS(\varkappa_2) = \cS(\varkappa_1 + \varkappa_2)\;,
\eeqn
implies that $\cS$ belongs to an Abelian (Cartan) subgroup of the $SU(1,1)$ group. 

The eigenvectors of the mass matrix ${\mathbf{e}}_\pm$, satisfying
\beqn
{\hat M} {\mathbf{e}}_\pm = M_\pm {\mathbf{e}}_\pm\;,
\eeqn
are equivalent to their bosonic analogues~\cite{Alexandre:2017foi}, see Sec.~\ref{sec:scalar}; namely,
\beqn
{\mathbf{e}}_+ =
\begin{pmatrix}
\phantom{-} \cosh \varkappa\\[1mm]
- \sinh \varkappa
\end{pmatrix},
\qquad
{\mathbf{e}}_- =
\begin{pmatrix}
\phantom{-} \sinh \varkappa\\[1mm]
- \cosh \varkappa
\end{pmatrix}. \quad
\label{eq_e:plus:minus}
\eeqn
Given that they are the eigenvectors of a non-Hermitian matrix, these vectors are not mutually orthogonal with respect to the ordinary Dirac inner product. Therefore, we introduce the auxiliary matrix
\beqn
A = \begin{pmatrix} 
\phantom{-} \cosh 2 \varkappa & \phantom{-} \sinh 2 \varkappa \\[1mm]
- \sinh 2 \varkappa & - \cosh 2 \varkappa
\end{pmatrix}\,,
\label{eq_A:matrix}
\eeqn
with $A \cdot A = \bbbone$. This allows us to identify a $\mathcal{APT}$ inner product that mirrors the scalar case in Sec.~\ref{sec:scalar}, where the matrix
\beqn
P \equiv P^{-1} = \sigma_3 \equiv
\begin{pmatrix} 
1 & 0 \\[1mm]
0 & - 1
\end{pmatrix}\,,
\label{eq_P:matrix}
\eeqn
is used to fulfil the analogous pseudo-Hermiticity condition
\beqn
P \cdot {\hat M} \cdot P = {\hat M}^{\sf T}
\eeqn
for the skew-symmetric mass matrix~\eq{eq_mass_matrix}. Here, the operator $\sf T$ denotes the matrix transpose.

In the next section, as was done for the scalar model in Sec.~\ref{sec:scalar} and the single Dirac model in Sec.~\ref{sec:oneflavour}, we will make the mass parameters of this two-flavour model coordinate dependent and gauge the corresponding similarity transformation~\eq{eq_similarity:transformation}. 

%%%%%%%%%%%%%%%%%%%%

\subsection{Local similarity transformation}
\label{sec:local}

To gauge the similarity group, we promote the global parameter $\varkappa$, entering the similarity matrix~\eq{eq_similarity:matrix}, to a local, spacetime-dependent quantity, $\varkappa = \varkappa(x)$. The similarity transformation becomes a local transformation $S = S(x) \equiv S[\varkappa(x)]$, which requires, in analogy with the usual gauge invariance, the appearance of a new vector matrix-valued similarity gauge field ${\cal C}^\mu$. 

The usual derivative is promoted to the covariant one, $\partial_\mu \to D_\mu$, with
\beqn
D_\mu = \bbbone \, \partial_\mu - {\cal C}_\mu \equiv \bbbone \, \partial_\mu + \sigma_1 C_\mu 
\equiv 
\begin{pmatrix}
\partial_\mu & C_\mu \\
C_\mu & \partial_\mu 
\end{pmatrix}
\,.
\label{eq_D_mu}
\eeqn
The similarity gauge field ${\cal C}_\mu \equiv - \sigma_1 C_\mu$ transforms under the local gauge similarity transformation as follows:
\beqn
{\cal C}_\mu \to \cS  {\cal C}_\mu \cS^{-1} - \cS  \partial_\mu \cS^{-1}\,.
\label{eq_C2:trans}
\eeqn
The vector field $C_\mu$ that enters the covariant derivative~\eq{eq_D_mu} transforms under the local similarity transformation~\eq{eq_C2:trans} as a $U(1)$ Cartan gauge field:
\beqn
C_\mu \to C_\mu + \partial_\mu \varkappa\,,
\label{eq_C2:trans:inf}
\eeqn
where we have used the relation $S^{-1} \partial_\mu S = - \sigma_1 \partial_\mu \varkappa$. 

The similarity-gauged fermionic model~\eq{eq_L} then acquires the following form:
\beqn
\cL_{\Psi,\mathcal{C}} = \bar \Psi \left(i \Dirac - {\hat M} \right) \Psi\,,
\label{eq_L:gauged}
\eeqn
where the superscript ``$C$'' indicates that the model is similarity-gauged. The covariant derivative is given in Eq.~\eq{eq_D_mu}, and $\Dirac = \gamma^\mu D_\mu$. Written explicitly, the Lagrangian~\eq{eq_L:gauged} is
\beqn
\cL_{\Psi,\mathcal{C}} & = & \sum_{a=1}^2 {\bar \psi}_a \left(i \dirac - m_a \right) \psi_a \nonumber \\
& & + {\bar \psi}_1 \left(m_5 + {\slashed C}\right) \psi_2 - {\bar \psi}_2 \left(m_5 - {\slashed C}\right) \psi_1\;,
\label{eq_L:gauged:explicit}
\eeqn
wherein we see that the similarity gauge field is associated with the off-diagonal terms.

The local similarity transformation can be summarised as follows
\beqn
\begin{array}{l}
{\mathrm{Local}} \\ 
{\mathrm{similarity}} \\
{\mathrm{transform}}
\end{array}
\!:\quad 
\left\{
\begin{array}{lcl}
\Psi(x) & \to & S(x) \Psi(x) \\ 
\bar \Psi(x) & \to & \bar \Psi(x) S^{-1}(x) \\
C_\mu(x) & \to & C_\mu(x) + \partial_\mu \varkappa(x)
\end{array}\;,
\right.
\qquad
\label{eq_loc_similarity}
\eeqn
where $S(x) \equiv  S[\varkappa(x)]$ is the local similarity transformation that depends on the arbitrary similarity parameter $\varkappa = \varkappa(x)$, and the $SU(1,1)$ matrix $S$ is given in Eq.~\eq{eq_similarity:matrix}. The similarity transformation law~\eq{eq_loc_similarity} is strikingly similar to a usual electromagnetic $U(1)$ gauge transformation --- cf.\ Eq.~\eq{eq_electromagnetic:gauge} below --- with the ``similarity field'' $C_\mu$ playing the role of the gauge field.

It is not difficult to realise that the coupling of the similarity gauge field $C_\mu$ to the fermion doublet in the action~\eq{eq_L:gauged} --- or, in the explicit form, in Eq.~\eq{eq_L:gauged:explicit} --- is given by the non-Hermitian term. This property implies that even in the absence of the off-diagonal non-Hermitian mass, viz.~$m_5 = 0$, the action~\eq{eq_L:gauged:explicit} corresponds to a non-Hermitian theory provided the vector similarity field is non-vanishing, i.e., $C_\mu \neq 0$.

%%%%%%%%%%%%%%%%%%%%

\subsection{Similarity gauge field and similarity current}

Before delving into the consequences of the presence of the similarity gauge field, it is appropriate to ask whether this field can be treated as an independent gauge field with its own kinetic term or if it must be considered as a non-dynamical background.

The existence of the similarity gauge invariance~\eq{eq_C2:trans:inf} of the fermionic action implies that the similarity gauge field~$C_\mu$ transforms as the usual $U(1)$ gauge field~\eq{eq_loc_similarity}, thus suggesting that the fermionic action~\eq{eq_L:gauged:explicit} can be supplemented with the gauge-invariant kinetic term for the gauge field
\beqn
\cL_{\Psi,\mathcal{C}} \overset{?}{\supset} - \frac{1}{4 g^2} f_{\mu\nu} f^{\mu\nu}, 
\qquad 
f_{\mu\nu} = \partial_\mu C_\nu - \partial_\nu C_\mu, \qquad
\label{eq_gauge:gS}
\eeqn
where $g$ is a new gauge coupling. This observation seems to support the idea that the gauge field $C_\mu$ can be quantised. However, as we show below, the appearance of the kinetic term~\eq{eq_gauge:gS} makes the theory inconsistent, thus forcing us to abandon the idea of a dynamical similarity gauge field and consider the field $C_\mu$ only as a background (classical) gauge field. 

The similarity gauge field couples to the similarity current \smash{$J^{\mathrm{(S)}}_\mu$} in the same way that the usual $U(1)$ photon gauge field couples to the electric current. The similarity current is given by a variation of the non-Hermitian matter action~\eq{eq_L:gauged} with respect to the similarity gauge field $C_\mu$ with the result
\beqn
J^{\mathrm{(S)}}_\mu = {\bar \Psi} \gamma^\mu \sigma^1 \Psi 
\equiv {\bar \psi}_1 \gamma^\mu \psi_2 + {\bar \psi}_2 \gamma^\mu \psi_1 \,.
\label{eq_J:S}
\eeqn

Using the classical equations of motion for non-Hermitian fermions~\eq{eq_L:gauged}
\beqs
\beqn
\left(i {\vec \Dirac} - {\hat M} \right) \Psi & = & 0\,, \label{eq_EQ_1}\\
{\bar \Psi} \left(i {\cev \Dirac} - {\hat M} \right) & = & 0\,,
\eeqn
\label{eq_EoM}
\eeqs
one can easily check that the similarity current, contrary to the electric current, is not conserved; namely,
\beqn
\partial^\mu J^{\mathrm{(S)}}_\mu = {\bar \Psi} \left[ 2 i m_5 \sigma_3 + (m_1 - m_2) \sigma_2 \right] \Psi\,.
\label{eq_d_JS}
\eeqn
Notice that this current is not conserved also in the Hermitian limit $m_5\to 0$.
The non-conservation property implies that the similarity gauge field $C_\mu$ must necessarily be made non-propagating so that the kinetic term~\eq{eq_gauge:gS} should not appear in the action. Otherwise, a variation of the action with respect to the gauge field would produce an inconsistent Maxwell-like equation \smash{$\partial^\nu f_{\mu\nu} = J^{\mathrm{(S)}}_\mu$} in which the right-hand-side would have a zero divergence while the left-hand-side would not. Notice that the conservation of the similarity current~\eq{eq_d_JS} is achieved in a trivial limit when and only when the mass matrix~\eq{eq_mass_matrix} becomes diagonal with equal eigenvalues. 

Thus, the gauge similarity field $C_\mu$ should be considered as parametrising some classical background, and could not provide, e.g., a candidate dark photon (for a review, see Ref.~\cite{FabbrichesiBook}). 

%%%%%%%%%%%%%%%%%%%%

\subsection{Constant physical masses and varying similarity background}

Proceeding as we did for the two-flavour scalar model in Sec.~\ref{sec:scalar} and the single Dirac fermion in Sec.~\ref{sec:oneflavour}, we consider the non-uniform non-Hermitian mass matrix
\beqn
{\hat M}(x) = 
\left(
\begin{array}{rr}
m_1(x) & \phantom{-} m_5(x) \\
- m_5(x) & m_2(x) 
\end{array}
\right)\,.
\label{eq_mass_matrix_xt}
\eeqn
The eigenvalues of this mass matrix can then be obtained with the help of a local similarity transformation:
\beqn
M_\pm(x) & = & M(x) \pm m(x)\,, \\ 
{\hat M}_{\rm diag}(x) & = & {\mathrm{diag}}\left( M_+(x), M_-(x) \right)\,,
\label{eq_mass:eigenvalues:3inh}
\eeqn
wherein we have used the parametrisation~\eq{eq_m_hat} with $M=M(x)$, $m=m(x)$, and $\varkappa = \varkappa(x)$.

In general, the physical masses~\eq{eq_mass:eigenvalues:3inh} are also spacetime-inhomogeneous quantities. Motivated by experimental and observational constraints on the spacetime variation of fundamental parameters, we again consider the case in which the physical masses are spacetime-independent.
This is implemented straightforwardly in the parametrisation~\eq{eq_m_hat} by setting the mass parameters $M$ and $m$ to be constant quantities ($M_0$ and $m_0$, respectively), thus endowing only the similarity scalar field $\varkappa$ with a coordinate dependence, i.e., $\varkappa = \varkappa(x)$. The mass matrix~\eq{eq_mass_matrix_xt} can then be written in the following form:
\beqn
{\hat M}(x) & = & \begin{pmatrix} M_0  & 0 \\ 0 & M_0  \end{pmatrix} 
\label{eq_mass_matrix_xt2} \\
& & 
+ \, m_0 \left(
\begin{array}{rr}
 \cosh 2 \varkappa(x) & \phantom{-} \sinh 2 \varkappa(x) \\
- \sinh 2 \varkappa(x) & - \cosh 2 \varkappa(x) 
\end{array}
\right)\nonumber \\
& \equiv & M_0 \, \bbbone + m_0 \big[ \sigma_3 \cosh 2 \varkappa(x) + i \sigma_2 \sinh 2 \varkappa(x)\big]\,.
\nonumber 
\eeqn
As we have already noticed, despite the spatial inhomogeneity of the mass matrix~\eq{eq_mass_matrix_xt2}, the eigenvalues [cf.~\eq{eq_mass:eigenvalues:2}]
\begin{equation}
M_0^{(\pm)} = M_0 \pm m_0
\end{equation}
are then constant quantities. Alternatively, in units of the entries of the mass matrix~\eq{eq_mass_matrix}, the parametrisation~\eq{eq_mass_matrix_xt2} implies that the following mass combinations are constant:
\beqn
m_1(x) + m_2(x) & = & 2 M_0\,, \\{}
[m_1(x) - m_2(x)]^2 - 4 m_5^2(x) & = & 4 m_0^2\,. 
\eeqn
Without loss of generality, we take $M_0 >0$ and $m_0 > 0$.

The non-Hermitian mass matrix~\eq{eq_mass_matrix_xt2} can be readily diagonalised, leading us to the Lagrangian
\beqn
\cL_{\Psi,\mathcal{C},{\rm diag}} = \bar \Psi \left(i {\hat\Dirac} - {\hat M}_{\rm diag}(x) \right) \Psi 
\equiv \bar \Psi {\hat K} \Psi \,,
\label{eq_L:Hermitian:local}
\eeqn
where 
\beqn
{\hat K} = 
\begin{pmatrix}
i \dirac - M_+ & - i \dirac \varkappa\\
- i \dirac \varkappa & i \dirac - M_-
\end{pmatrix}\,,
\label{eq_K_1}
\eeqn
and wherein we recognise the similarity field~$C_\mu = \partial_\mu \varkappa(x)$. The structure~\eq{eq_K_1} represents a non-Hermitian operator in which the violation of Hermiticity is determined by the magnitude of the off-diagonal terms. 

In order to proceed further, it is convenient to work in a linear approximation with a slowly varying scalar similarity parameter
\beqn
\varkappa(x) = \varkappa_0 + C_\mu x^\mu\;,
\label{eq_kappa_x}
\eeqn
where $\varkappa_0$ is a background (coordinate-independent) quantity and the linear variation of the scalar similarity field is given by the vector $C^\mu$.

In terms of the masses~\eq{eq_mass_matrix_xt2}, the inhomogeneous background~\eq{eq_kappa_x} corresponds to the following mass matrix:
\beqn
{\hat M}(x) & = & \left(
\begin{array}{rr} M_0 + m_0 \cosh 2 \varkappa_0 & 
m_0 \sinh 2 \varkappa_0 \\ - m_0 \sinh 2 \varkappa_0 & M_0 - m_0 \cosh 2 \varkappa_0 
\end{array} \right)
\nonumber\\
& & 
+ \, m_0 \left(
\begin{array}{rr}
 \sinh 2 \varkappa_0 & \phantom{-} \cosh 2 \varkappa_0 \\
- \cosh 2 \varkappa_0 & - \sinh 2 \varkappa_0
\end{array}
\right)  C_\mu x^\mu + O(x^2)\;.\nonumber\\
\label{eq_mass_matrix_xt3} 
\eeqn
We will work mostly with a slight non-Hermitian perturbation of Hermitian theories corresponding to $\varkappa_0 = 0$ because the constant (spacetime-independent) non-Hermitian background can always be reduced to a Hermitian theory by an inverse similarity transformation. In this case, the mass matrix~\eq{eq_mass_matrix_xt3} simplifies further to
\beqn
{\hat M}(x) & = & \left(
\begin{array}{rr} M_0 + m_0 & 
- m_0 \, C_\mu x^\mu\\[1mm] m_0 \, C_\mu x^\mu & M_0 - m_0 
\end{array} \right) + O(x^2)\,.
\label{eq_mass_matrix_xt4}
\eeqn

In the phenomenologically interesting limit of weak inhomogeneity, one gets from Eqs.~\eq{eq_mass_matrix} and~\eq{eq_mass_matrix_xt3} the following relation of the mass matrix elements and the similarity gauge field:
\begin{align}
    C_\mu = - \frac{\partial }{\partial x^\mu} \frac{m_5}{m} + \dots \equiv - \frac{2}{M_+ - M_-} \frac{\partial m_5}{\partial x^\mu} + \dots \,.
    \label{eq_Cmu_partial}
\end{align}
This is to say that we assume the background field to develop a small, weakly inhomogeneous off-diagonal non-Hermitian mass, which, in turn, can be treated as the emergence of a weak similarity gauge field~$C_\mu$. This is connected to the inhomogeneity of the off-diagonal mass via Eq.~\eq{eq_Cmu_partial} with $|C_{\mu} x^{\mu}| \ll 1$ at the length scale of the inhomogeneities of $C_\mu$. Moreover, we  assume that the strength of the similarity field $C_\mu$ is much smaller than the mean values $M_0 = (M_{+,0} + M_{-,0})/2$ and the splitting $m_0 = (M_{+,0} - M_{-,0})/2$ of the physical masses $M_{\pm,0}$ at $C_\mu = 0$, given by Eqs.~\eq{eq_mass:eigenvalues:2}, \eq{eq_mass:eigenvalues:2} and \eq{eq_mass:eigenvalues:3}, i.e., 
\begin{align}
    |C_0|, \, |{\bs C}| \ll M_0,\,  m_0\,.
    \label{eq_C_small}
\end{align}

The operator~\eq{eq_K_1}, applied to a spinor plane wave eigenstate $\Psi(x) = e^{-i p \cdot x} \Psi_0$, with a constant spinor doublet $\Psi_0$, becomes 
\beqn
{\hat K} = 
\begin{pmatrix}
\slashed{p} + M_+ & - i {\slashed C} \\
- i {\slashed C} & \slashed{p} + M_-
\end{pmatrix}\;.
\label{eq_K_2}
\eeqn
The condition for the eigenvalues~\eq{eq_EQ_1} transforms to the compatibility equation $\det {\hat K} = 0$, which implies, in turn, the following relation:
\begin{align}
        & (p^2 - M_+^2 + C^2) (p^2 - M_-^2 + C^2) \label{eq_omega_1}\\
        & \hskip 10mm - C^2 \bigl[4 p^2 - (M_+ + M_-)^2 \bigr] + 4 (C \cdot p)^2 = 0\,.
\nonumber 
\end{align}

Equation~\eq{eq_omega_1} determines the energy dispersion relations for the fermions. In the absence of the similarity gauge background, i.e., $C^\mu = 0$, this equation gives us two standard excitation branches 
\begin{align}
    \omega^2_{\pm,\bs{p}} = {\bs p}^2+M_\pm^2, \qquad\ {\text{for}}\quad C^\mu = 0\,,
\end{align}
where the physical real-valued masses $M_\pm$ are given in Eq.~\eq{eq_mass:eigenvalues}. The positive-definite inner product for these mass eigenstates (generalising the construction in Sec.~\ref{sec:globalcase}) is described in Appendix~\ref{sec:innerprod}.

%%%%%%%%%%%%%%%%%%%%

\section{Inhomogeneous similarity background: energy dispersions}
\label{sec:dispersion}

As for the scalar case in Sec.~\ref{sec:scalar}, the energy eigenvalues are roots of an algebraic equation of the 4th order, given by Eq.~\eq{eq_omega_1}. It is, therefore, convenient again to consider the cases of purely timelike or purely spacelike similarity gauge fields.

%%%%%%%%%%%%%%%%%%%%

\subsection{Temporal similarity field}

A time-varying and spatially homogeneous mass matrix provides us with a strictly temporal background similarity field $C^\mu = (C^0, {\bs 0})$, with $C^0 \equiv C_0$. Equation~\eq{eq_omega_1} then gives us the following energy dispersion relations:
\begin{align} \label{eq_omega_C0}
    \omega_{\pm,\bs{p}}^2 &=  M_0^2 + m_0^2 + {\bs p}^2 - C_0^2  \nonumber\\
    & \phantom{=}\pm 2 \sqrt{M_0^2 (m_0^2 - C_0^2) - C_0^2 {\bs p}^2} \nonumber \\
    & \hskip 25mm {\text{for}}\ C^\mu = (C^0,{\bs 0})\;.
\end{align}
We describe the solutions in terms of the unperturbed, coordinate-independent masses $M_0$ and $m_0$ in the absence of the background similarity field ($C^\mu = 0$). Their relation to the unperturbed mass matrix~\eq{eq_mass_matrix} can be read from Eq.~\eq{eq_mass:eigenvalues:3}.

The energy dispersion relation~\eq{eq_omega_C0} has the following notable features:
\vskip 1mm
\paragraph*{\bf -- Mass shift.}
The masses $M_{\pm,C} \equiv \omega_{\pm} ({\bs p} = 0)$ are affected by the presence of the similarity gauge field, as is readily visible from Eq.~\eq{eq_omega_C0}. Specifically,
\begin{align} \label{eq_mass_C0}
   M_{\pm,C} & = \left(M_0^2 + m_0^2 - C_0^2  \pm 2 M_0 \sqrt{m_0^2 - C_0^2} \right)^{1/2} \nonumber\\
   & = M_{\pm,0} \mp \frac{C_0^2}{M_{+,0} - M_{-,0}} + \dots\,,
\end{align}
where the ellipsis denotes $O\bigl(C_0^4\bigr)$ terms and the unperturbed masses are $M_{\pm,0} = M_0 \pm m_0$ in consistency with Eq.~\eq{eq_mass:eigenvalues:2}. In the leading, quadratic order, the weak field $C_0$ slightly contributes to the shift between the $M_{\pm,C}$ masses, while leaving their mean value unchanged.

\vskip 1mm
\paragraph*{\bf -- High-momentum instability.} The dispersion relation~\eq{eq_omega_C0} develops an imaginary part at a certain spatial momentum $p = |{\bs p}|$, which is restricted by the higher cutoff $p_{c}^{(t)}$, defined via
\begin{align}
    p > p^{(t)}_{c} = M_0 \sqrt{\frac{m_0^2}{C_0^2} - 1}\;.
    \label{eq_p_max_C0}
\end{align}
Given the weakness of the similarity gauge field~\eq{eq_C_small}, the limiting momentum at which the particle propagation becomes unstable is much higher than the mean mass of the particles, viz.~$p_{\mathrm{max}}^{(t)} \gg M_0$.

\begin{figure}[!ht]
\begin{center}
\includegraphics[width=0.45\textwidth,clip=true]{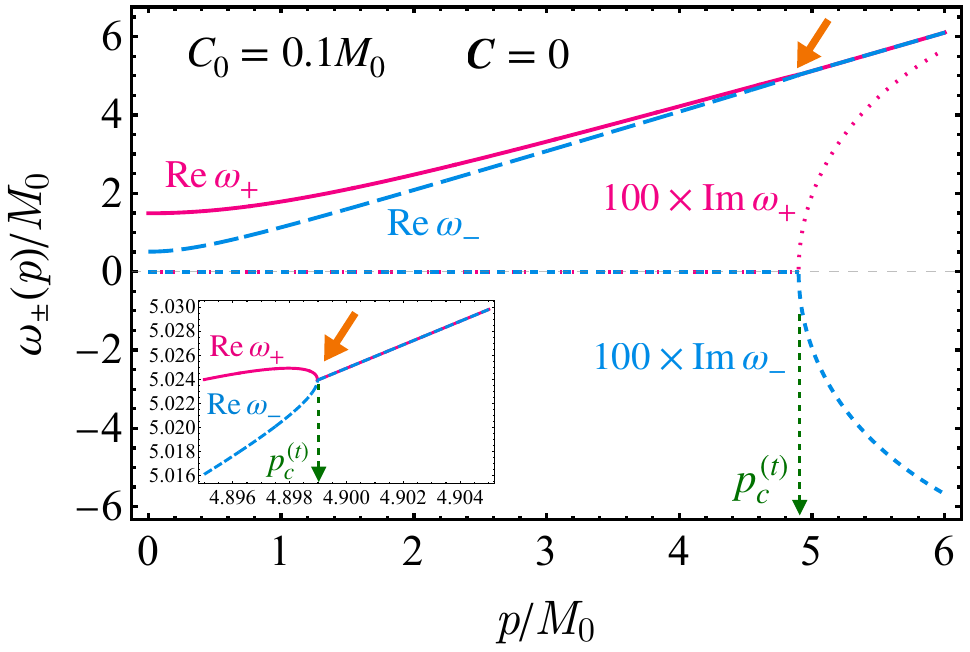}
\end{center}
\caption{The real and imaginary parts of the energy dispersions $\omega_\pm$ for the timelike similarity field $C_0 \neq 0$ and ${\bs C} = 0$ are shown for the illustrative set of parameters $m_0 = M_0/2$ and $C_0 = M_0/10$. The imaginary parts, multiplied by a factor of 100 to increase their visibility, appear at the critical momentum~\eq{eq_p_max_C0} \smash{$p^{(t)}_{c} \simeq 4.90$}, identified with the green arrow. The inset shows the real part of the dispersions close to the critical point \smash{$p \simeq p^{(t)}_{c}$} (at the location indicated with the orange arrows).}
\label{fig_omega_C0}
\end{figure}

At large momentum $p \gg p_{c}^{(t)}$, one gets the following asymptotic form of the energy dispersion:
\begin{align}
    \omega_{\pm, p} = p \pm i |C_0| + \frac{M_0^2 + m_0^2}{2 p} + O\bigl(p^{-2}\bigr)\;.
    \label{eq_omega_pm_C0}
\end{align}
In the selected parameter region~\eq{eq_C_small}, the instability given by the imaginary term in the dispersion~\eq{eq_omega_pm_C0} is very small. Examples of the dispersion relation are illustrated in Fig.~\ref{fig_omega_C0} for a suitable set of parameters. Notice that, according to Eqs.~\eq{eq_mass_C0} and \eq{eq_omega_pm_C0}, in the unstable region, \smash{$p > p^{(t)}_{c}$}, the particle with the lower mass $M_-$ has a diffusive nature --- the mode decays since $\mathrm{Re}\, \omega_- < 0$ --- whereas the higher-mass $M_+$ particle is unstable --- the mode grows because $\mathrm{Re}\, \omega_+ > 0$. 

Given the similarities in the form of their dispersion relations, the instability of the fermion modes at high momentum is analogous to the instability found for the two-scalar model in Sec.~\ref{sec:scalar}, see Ref.~\cite{Chernodub:2021waz}.  Even so, from a general point of view, it is an unexpected feature of the presence of the timelike similarity field.
A similar bound to Eq.~\eq{eq_p_max_C0} is obtained for the counterpart bosonic model~\cite{Chernodub:2021waz}, see Sec.~\ref{sec:scalar}. As noted earlier for both the scalar and single Dirac fermion models, these instabilities represent momentum-dependent exceptional points, beyond which we obtain complex conjugate pairs of eigenfrequencies.

\vskip 1mm
\paragraph*{\bf -- Superluminal propagation.} In the unstable region, the group propagation velocities of both modes 
\begin{align}
    {\bs v}_{\pm,\bs{p}} = \frac{\partial \omega_{\pm,\bs p}}{\partial {\bs p}}\;, 
    \label{eq_v_pm}
\end{align}
defined by the slope of the real part of the dispersion~\eq{eq_omega_pm_C0}, always remain smaller than the speed of light, i.e., $|{\bs v}_{\pm,\bs{p}}| < 1$. However, for high momenta of the order of $p^{(t)}_{c}$ --- with the condition that they are still lower than the instability threshold, \smash{$p < p^{(t)}_{c}$} --- the velocity of particle propagation~\eq{eq_v_pm}, for both ``$\pm$'' modes, exceeds the speed of light in a certain region of parameter space. This property is apparent from the inset of Fig.~\ref{fig_omega_C0}, which zooms in on the dispersions~\eq{eq_omega_C0} in a small region around the critical momentum~\eq{eq_p_max_C0}. The cusps originating from the square root of the dispersions~\eq{eq_omega_C0}, show that the velocity of both $\omega_\pm$ modes becomes singular at \smash{$p = p^{(t)}_{c}$}.

The superluminal propagation is an unanticipated property of the non-Hermitian model. This peculiar feature suggests that fast-moving particles, coupled with other dynamical fields like the photon field, would emit Cherenkov radiation. This radiation, in turn, would act as a decelerating force on the particles, effectively impeding their advancement toward the point of instability for all modes.

\begin{figure}[!ht]
\begin{center}
\includegraphics[width=0.475\textwidth,clip=true]{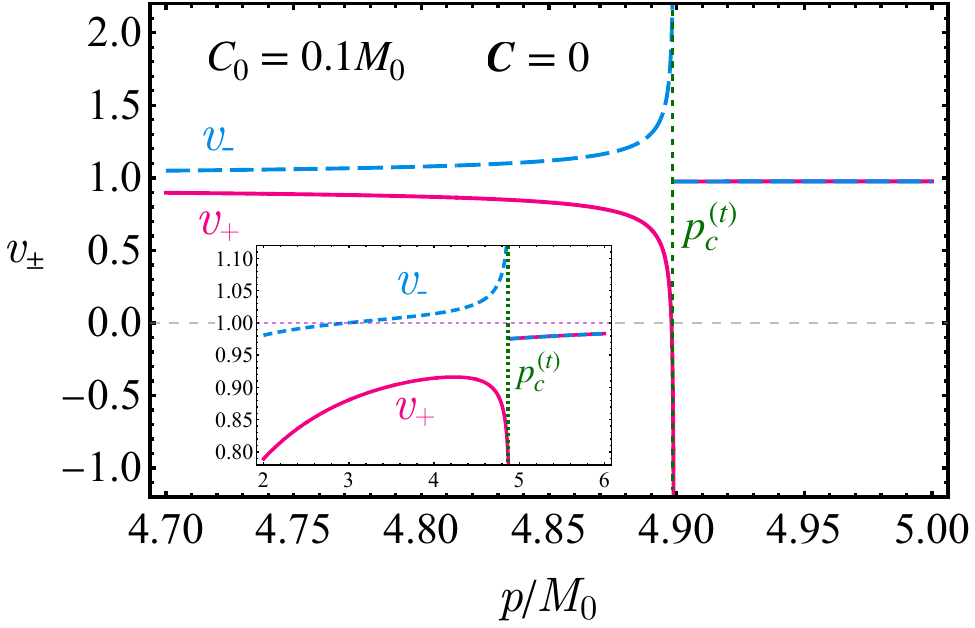}
\end{center}
\caption{Velocities of the ``$\pm$'' modes of Fig.~\ref{fig_omega_C0} for the timelike similarity field. The inset shows a wider region in the momentum $p$ space narrowed in the $v$-axis around the speed of light, $v = 1$.}
\label{fig_v_C0}
\end{figure}

The group velocities for both ``$\pm$'' modes are illustrated in Fig.~\ref{fig_v_C0} in which the parameters of Fig.~\ref{fig_omega_C0} are adopted. For the ``$-$''-mode, the superluminal propagation arises at high momenta \smash{$p > p^{\mathrm{(t,-)}}_{\mathrm{SL}}$}, where
\begin{align}
	p^{{(t,\pm)}}_{\mathrm{SL}} = & \frac{M_0}{|C_0|} \frac{\sqrt{m_0^2 - C_0^2}}{M_0^2 + m_0^2} \biggl(C_0^2 \left(M_0^2-m_0^2\right) + M_0^4 + m_0^4 \nonumber \\
    & \pm 2 m_0 M_0 \sqrt{(m_0^2 - C_0^2) (M_0^2 + C_0^2)} 
    \biggr)^{1/2}\,.
	\label{eq_p_t_SL}
\end{align}
As was the case for the scalar model, the superluminality (``SL'') threshold momentum~\eq{eq_p_t_SL} for the ``$-$''-mode
is of the order of, but noticeably lower than, the critical momentum~\eq{eq_p_max_C0} at which the instability sets in. In the lower-momentum region, \smash{$p < p^{{(t)}}_{\mathrm{SL}}$}, the propagation of this mode is characterised by subluminal velocities. 

\vskip 1mm
\paragraph*{\bf -- Stopped propagation.} The ``+''-mode has a richer structure. As the momentum $p$ increases, the group velocity of this mode reaches a maximum in the subluminal region (which is about 90\% of the speed of light for our choice of parameters) before the velocity of the particle then drops again. When the momentum reaches a particular value,
\begin{align}
	p_{\mathrm{stop}}^{(t)} = \frac{\sqrt{M_0^2 m_0^2 - M_0^2 C_0^2 - C_0^4}}{|C_0|}\,,
	\label{eq_p_st}
\end{align}
the ``$+$''-mode stops propagating as the group velocity vanishes,  i.e., 
\begin{align}
	v_+\bigl(p_{\mathrm{stop}}^{(t)}\bigr) = 0\,.
	\label{eq_stop}
\end{align}
Notice that for weak similarity fields~\eq{eq_C_small}, the stopping momentum~\eq{eq_p_st} is very close to the instability point~\eq{eq_p_max_C0}. 

As the momentum increases further, the group (negative) velocity exceeds the speed of light and diverges. The superluminal thresholds for both ``$\pm$'' modes can be expressed via the following single expression:
\begin{align}
	v_\pm \bigl(p^{{(t,\pm)}}_{\mathrm{SL}}\bigr) = \mp 1\,,
	\label{eq_v_SL}
\end{align}
where the velocities $v_\pm$ are expressed in Eq.~\eq{eq_v_pm} via the corresponding frequencies~\eq{eq_omega_C0}. 

\vskip 1mm
\paragraph*{\bf -- Negative group velocity.}

As the momentum increases above the ``stopping point'' $p = p^{{(t,+)}}_{\mathrm{SL}}$, the ``$+$''-mode starts to propagate in a backward direction because the group velocity takes a negative value, i.e.,
\begin{align}
	 v_+(p) < 0 \qquad \text{for} \quad p^{{(t,+)}}_{\mathrm{SL}} < p < p^{(t)}_{c}\,.
	\label{eq_backwards}
\end{align}
This effect persists in a narrow region close to the critical momentum~\eq{eq_p_max_C0}, beyond which both ``$\pm$''-modes develop a complex part.

The emergence of a negative group velocity for wave propagation is an interesting phenomenon, which often appears in optics~\cite{Brillouin2013wave}. The negative group velocities, as well as the superluminal propagation, is also a characteristic feature of media with anomalous dispersion~\cite{Garrett1970}, which we discuss below. 

\vskip 1mm
\paragraph*{\bf -- Anomalous dispersion.} The non-Hermitian two-fermion model in the background of a nonvanishing similarity gauge field also features anomalous dispersion. One can readily check that the ``$+$''-mode of our dispersion relation~\eq{eq_omega_C0} is always normal with $\partial n_+ / \partial p > 0$ in the whole range of frequencies. However, the ``$-$''-mode exhibits anomalous dispersion in a wide region of the phase diagram.

It is worth mentioning that properties such as superluminality (with the group velocity exceeding the speed of light) and negative group velocity (which is opposite to the wavevector) often appear in homogeneous media characterised by anomalous dispersion~\eq{eq_anomalous}. The consequences of anomalous dispersion on the shape of a Gaussian light pulse as it propagates through such a medium were studied in Ref.~\cite{Garrett1970}, where it was found that the pulse remains of the Gaussian shape with the peak moving in space with a velocity determined by the classical group-velocity expression~\eq{eq_v_gr_n}. 

Amusingly, in the anomalously dispersive medium, the classical group velocity can become greater than the velocity of light in a vacuum or negative, provided the medium has an absorption line near the optical frequencies of the waves that make up the pulse. The absorption property is a characteristic of a non-Hermitian system, which points out that the non-Hermiticity is responsible for the unification of all three phenomena. 

These would-be contradictory statements made in the classical context do not mean that the system necessarily violates causality. The mechanism beyond the superluminal propagation in anomalous dispersive media is associated with the pulse shape distortion, even though in their studies, the pulse does not appear to be visually distorted~\cite{Garrett1970}. Causality is also not violated in such an anomalous medium since the absorption destroys the relation between the group velocity and the velocity of energy propagation. While we expect that a similar phenomenon can also happen in the present non-Hermitian model, we leave a more detailed study for future work. 

\begin{figure*}[!htb]
\begin{center}
\includegraphics[width=0.95\textwidth,clip=true]{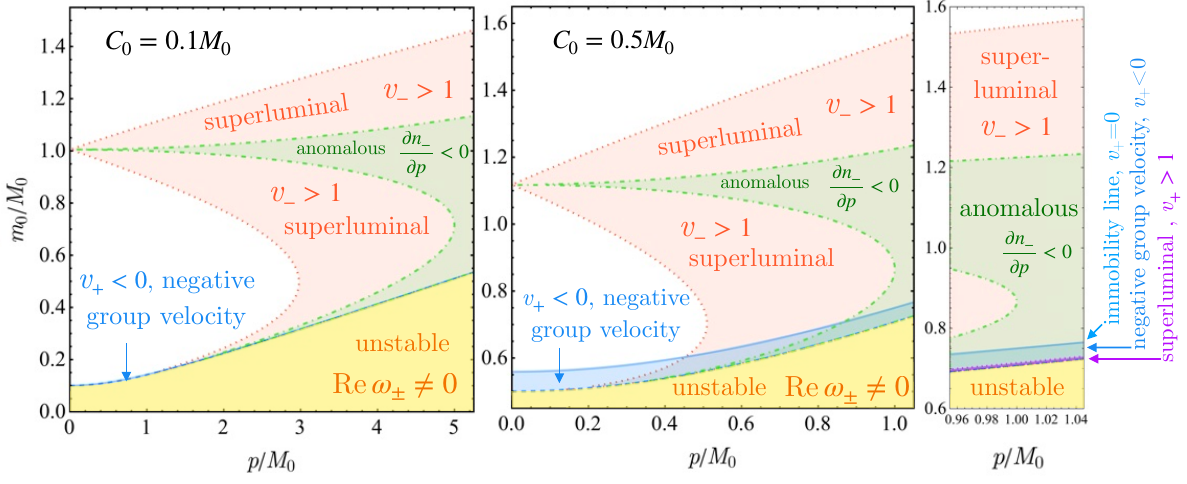}  
\end{center}
\caption{The regions in the $(p,m_0)$ plane at $C_0 = 0.1 M_0$ (left) and $C_0 = 0.5 M_5$ (middle) of:\ the high-momentum instability (yellow for both modes), superluminal propagation (red for ``$-$''-mode and magenta for ``$+$''-mode), anomalous dispersion of the ``$-$''-mode (green), the line of immobility of the ``$+$''-mode (blue), the negative group velocity of the ``$+$''-mode (blue). A zoom-in on a region around $p = M_0$ for $C_0 = 0.5 M_5$ (right) also shows the region where the ``$+$''-mode becomes superluminal with negative group velocity (magenta). The blue and yellow regions are separated by the critical momentum~\eq{eq_p_max_C0}.}
\label{fig_phase_diagrams}
\end{figure*}

We finish this subsection with Fig.~\ref{fig_phase_diagrams}, which illustrates all the described particularities of the propagation of the ``$\pm$''-fermionic modes for two inhomogeneity parameters corresponding to a relatively weak ($C_0 = 0.1 M_0$) and moderate ($C_0 = 0.5 M_0$) similarity fields. 

%%%%%%%%%%%%%%%%%%%%

\subsection{Spatial similarity field}

A strictly spatial perturbation of the mass matrix is equivalent to introducing the similarity field $C^\mu = (0,{\bs C})$. In this case, Eq.~\eq{eq_omega_1} gives us the following dispersion:
\begin{align} \label{eq_omega_C}
    \omega_{\pm,\bs{p}}^2 = & M_0^2 + m_0^2 + {\bs p}^2 - {\bs C}^2\nonumber \\
    & \pm 2 \sqrt{m_0^2 (M_0^2 - {\bs C}^2) - ({\bs C} \cdot {\bs p})^2}  \nonumber \\
    & \hskip 25mm {\text{for}}\quad C^\mu = (0,{\bs C})\,,
\end{align}
which is naturally anisotropic due to the presence of the spatial vector ${\bs C}$. Notice again that the fermionic dispersions are functionally similar, up to redefinitions of the masses, to their bosonic counterparts discussed in Sec.~\ref{sec:scalar}, see Ref.~\cite{Chernodub:2021waz}.

The resemblance of the dispersions~\eq{eq_omega_C} and Eq.~\eq{eq_omega_C0} implies that the spacelike similarity field has exactly the same features as the timelike field considered earlier. In particular, the shift of the fermion masses in the presence of the spatial similarity field is given by Eq.~\eq{eq_mass_C0} for the temporal similarity field with the replacements $m_0 \leftrightarrow M_0$ and $C_0 \to C_3$. The same statement also applies to the critical momenta~\eqref{eq_p_max_C0}, \eqref{eq_p_t_SL}, and \eqref{eq_p_st} that characterise the upper bound on stability modes, the superluminal momentum threshold, and the stopping momentum, respectively.

The high-momentum instability appears if the momentum of a fermion tangential to the similarity field ${\bs C}$ exceeds the following threshold:
\begin{align}
    |p_3| > p^{(s)}_{\mathrm{max}} = m_0 \sqrt{\frac{M_0^2}{C_3^2} - 1}\;.
    \label{eq_p_max_C}
\end{align}
Here, we took ${\bs C}$ along the third axis, $C^\mu = \delta^{\mu3} C_3$.  An increasing momentum in the direction normal to the similarity field ${\bs C}$ does not lead to instabilities. 

\begin{figure*}[!htb]
\begin{center}
\begin{tabular}{cc}
\includegraphics[width=0.45\textwidth,clip=true]{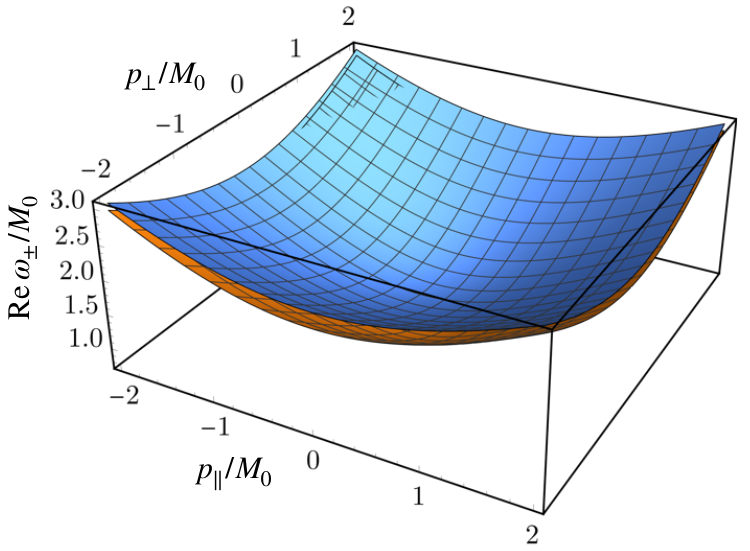} &  \includegraphics[width=0.45\textwidth,clip=true]{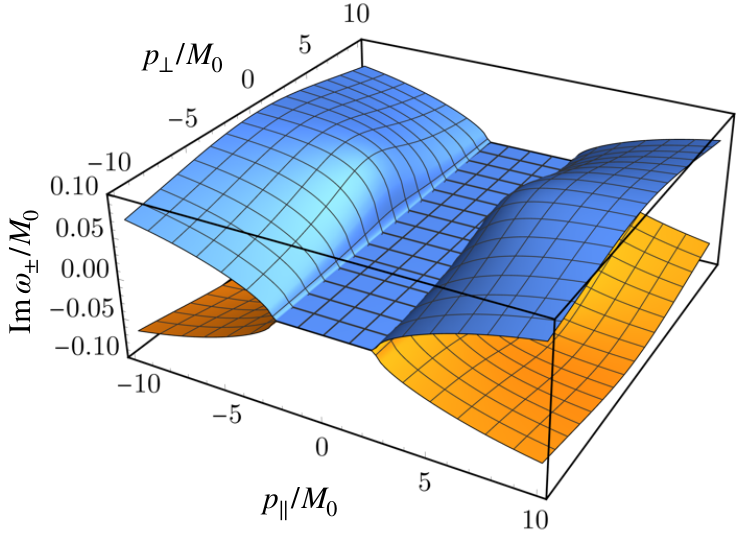} \\
\end{tabular}
\end{center}
\caption{The real (left) and imaginary (right) parts of the energy dispersions $\omega_\pm$ for the spacelike similarity field $C^\mu \equiv (0,0,0,C_3) \neq 0$ for the same set of parameters as in Fig.~\ref{fig_omega_C}. The imaginary parts develop when the longitudinal momentum $p_\| \equiv p_3$ exceeds the critical momentum \smash{$p^{(s)}_{\mathrm {max}}$} [see Eq.~\eq{eq_p_max_C}]. The plots are shown in the $(p_\|,p_\perp)$ plane, where \smash{$p_\perp = \pm \sqrt{p_1^2 + p_2^2}$} is the transverse momentum.}
\label{fig_omega_C}
\end{figure*}

At large momentum $p_\| \equiv p_3 \gg p^{(s)}_{\mathrm{max}}$, one gets the asymptotic form of the energy dispersion similar to Eq.~\eq{eq_omega_pm_C0}, 
\begin{align}
    \omega_{\pm} (p_\perp,p_3) = p \pm i |C_3| + \frac{M_0^2 + m_0^2 + p^2_\perp}{2 p_3} + O\bigl(p^{-2}_3\bigr)\,,
    \label{eq_omega_pm_C}
\end{align}
whereas in the limit $p_\perp = \sqrt{p_1^2 + p_2^2} \to \infty$ (and in the stable region \smash{$| p_\| | < p^{(s)}_{\mathrm{max}}$}), we arrive to the expected dispersion $\omega_{\pm} (p_\perp,p_3) = p + O(p^{-1})$, which does not contain an imaginary part. The behaviour of the real and imaginary parts of the dispersion relation~\eq{eq_omega_C} for the spacelike similarity field are illustrated in Fig.~\ref{fig_omega_C}.

The superluminal propagation, negative group velocity and anomalous dispersion are also evident from the similarity of the dispersions of fermions in the background of the spatial~\eq{eq_omega_C} and temporal~\eq{eq_omega_C0} similarity fields. Both of them contain a square-root term, which leads to the characteristic cusps of the dispersion at the critical momentum, which, in turn, implies the divergent group velocity of the modes. 

Interestingly, the dispersion relations for the two-flavour fermionic non-Hermitian model with temporal~\eq{eq_omega_C0} and spatial~\eqref{eq_omega_C} similarity gauge fields coincide with the corresponding dispersion relations for the non-Hermitian model with the doublet scalar field~\eq{eq_omega_scalar_cases} with obvious redefinitions of the critical momenta. Therefore, the phase diagrams for both models share the same features represented in Fig.~\ref{fig_phase_diagrams}. 

Finally, it is appropriate to ask whether these unusual features of the non-Hermitian model appear due to the non-Hermiticity or if they can also be found in a Hermitian version of the model. The model can be turned to its Hermitian version by making the off-diagonal term in the mass matrix~\eq{eq_mass_matrix_xt} a purely imaginary quantity $m_5 \to i m_5$ (with a real non-Hermitian mass $m_5$). This change implies, according to Eq.~\eq{eq_mass_matrix_xt2}, that the scalar similarity parameter $\varkappa$ also becomes a purely imaginary quantity, which also provokes, following Eq.~\eq{eq_kappa_x}, the corresponding replacement of the similarity gauge field, $C_\mu \to i A_\mu$, where $A_\mu$ is a real-valued vector field. Thus, in the Hermitian version of the model, the dispersion relations are given by Eqs.~\eq{eq_omega_C0} and \eq{eq_omega_C} with the replacement $C^\mu \to i C^\mu$. One can readily see that the Hermitian version of the model has an unremarkable monotonically rising behaviour as a function of momenta, devoid of any exotic properties. 

%%%%%%%%%%%%%%%%%%%%

\section{Instability, IR/UV mixing and phenomenology}
\label{sec:pheno}

The models described exhibit momentum-dependent exceptional points. In the case of the two-scalar and two-fermion models, the low momentum modes reside in the $\PT$-unbroken domain such that the low-lying physical excitations have a real-valued spectrum and are stable. On the other hand, at sufficiently high momenta, the modes reside in a $\mathcal{PT}$-broken regime. The lower-mass branch $\omega_-$ develops a negative imaginary part, implying that the amplitude of the mode decays. The higher-mass branch $\omega_+$ develops a positive imaginary part signalling the growth of this mode. 

The spacetime inhomogeneity of the mass matrix in realistic models is expected to result from the inhomogeneity of an underlying scalar condensate. In the fermion case, this condensate would couple to the fermion bilinear via a Yukawa term and contribute to the fermion mass terms. The instabilities of the high-momentum fermion modes can, therefore, be understood as a result of the interaction between the inhomogeneous scalar condensate and a propagating fermionic particle:\ the particle scatters inelastically at the condensate creating or absorbing scalar quanta. One can expect that the fluxes of highly energetic fermionic particles propagating with momenta above the thresholds~\eq{eq_p_max_C0} and \eq{eq_p_max_C} through the scalar condensate will result in the homogenisation of the condensate. The translationally invariant condensate, corresponding to the true ground state of the model, would then be the ultimate outcome of these interactions. In our paper, we do not consider this homogenising backreaction of the propagating fermionic modes on the non-Hermitian background. 

For phenomenologically relevant weak similarity fields~\eq{eq_C_small}, the stability zones for the two-flavour fermion model, defined for the temporal~\eq{eq_p_max_C0} and spatial~\eq{eq_p_max_C} cases, determine the single critical momentum
\begin{align}
	p_{c} = \frac{M_0 m_0}{C}\;,
	\label{eq_pc}
\end{align}
where $C > 0$ stands for either $C_0$ or $|\bs C|$ and we have neglected the subleading terms in Eq.~\eq{eq_pc}. The above equation implies that a weak similarity field $C^\mu$, corresponding to a slowly varying mass matrix and a low (IR) momentum, leads to an instability of the propagating modes at high (UV) momentum. This IR/UV mixing was found in Ref.~\cite{Chernodub:2021waz}, and it is a particular feature of these non-Hermitian field theories.

An estimation of the critical momentum~\eq{eq_pc} can be made with the help of Eq.~\eq{eq_Cmu_partial}:
\begin{align}
	p_{c} = \frac{1}{8} (M_+ + M_-) (M_+ - M_-)^2 
            \biggl(\frac{\partial m_5}{\partial x^\mu}\biggr)^{-1}\,,
	\label{eq_pc_2}
\end{align}
where we have omitted subleading corrections of the order $O\bigl((\partial m_5/\partial x^\mu)/m_{1,2}^2\bigr)$. For this, let us use the physical lepton masses $M_- = m_e \simeq 0.5\,\mathrm{MeV}$ and $M_+ 
 = m_\mu \simeq 105.7\,\mathrm{MeV}$. (We note that if, indeed, we were to associate the fermion doublet with electrically charged degrees of freedom, conservation of the electromagnetic current for the non-Hermitian theory would require the components of the fermionic doublet to carry equal electric charge, see Appendix~\ref{sec:emcharges}.) If we assume that the non-Hermitian mass $m_5$ varies by one MeV at the distance of one meter, the critical momentum~\eq{eq_pc_2} then acquires the value 
\begin{align}
      p_c \simeq 1.6 \times 10^{28}\,\mathrm{MeV}\,,
    \label{eq_pc_estimation}
\end{align}
For these conditions, the similarity field~\eq{eq_Cmu_partial} is extremely small, with
\begin{align}
    |C| \simeq 3.8 \times 10^{-15}\,\mathrm{MeV}\,,
    \label{eq_C_mu_estimation}
\end{align}
so that the assumption of the weakness of the similarity field is satisfied~\eq{eq_C_small}. In other words, for the negligibly tiny inhomogeneity of the non-Hermitian masses encoded in the experimentally unobservable value of the similarity field~\eq{eq_C_mu_estimation}, the fermionic modes become unstable as soon as the momentum of the fermionic particle exceeds the critical value~\eq{eq_pc_estimation}. 

It is important to stress again that it is possible to vary the non-Hermitian mass parameters without affecting the physical masses $M_\pm$. Therefore, even if the value of these parameters vary from one spacetime region to another, the physical masses remain the same within a tiny, experimentally inaccessible correction of the order of the strength of the similarity vector field~\eq{eq_C_mu_estimation}. 

%%%%%%%%%%%%%%%%%%%%

\section{Conclusions}
\label{sec:conc}

In this article, we have extended the proposal of Ref.~\cite{Chernodub:2021waz} to non-Hermitian fermionic theories with local Lagrangian parameters, namely the parameters of the mass matrix. This leads automatically to the appearance of an associated vector field, the ``similarity gauge field'' $C^\mu$, which acts as a new connection in the space of similar non-Hermitian theories. 

We have argued that the similarity gauge field $C^\mu$ cannot be a dynamical propagating field similar, e.g., to the electromagnetic gauge field. However, the similarity field can still appear in the model as a non-dynamical, background field.

The spacetime-dependent contributions to the mass matrix can be viewed as a result of the inhomogeneity of a condensate of a scalar field coupled to the fermion doublet via, e.g., a Yukawa coupling. Such contributions can be of either Hermitian or non-Hermitian nature. Assuming a weak inhomogeneity of the scalar background, one can show, following the bosonic case studied in detail in Ref.~\cite{Chernodub:2021waz}, that inhomogeneity of a Hermitian mass matrix leads to a rather trivial effect, resulting in a shift in (or redefinition of) the fermionic masses. On the other hand, going beyond Ref.~\cite{Chernodub:2021waz}, we have shown that the inhomogeneity of a non-Hermitian mass mixing matrix leads to nontrivial effects both for two-flavour scalar and two-flavour fermion models:\ anomalous dispersion, superluminality and instabilities above certain high-momentum (UV) thresholds, determined by the IR scale of the inhomogeneity. More precisely, the longer the wavelength of the inhomogeneity, the higher the critical value of the momentum above which the instability arises. This IR/UV mixing effect was observed for the scalar case in Ref.~\cite{Chernodub:2021waz}. Thus, weak variations of the mass parameters on cosmological scales would correspond to instabilities at scales beyond current experimental or observational reach.

An unexpected feature of the two-flavour models is that the group velocity of the particle propagation exceeds the speed of light at sufficiently high momenta, which are of the same (high) magnitude as the critical momenta that mark the onset of the instability. This feature implies that the particles, depending on their couplings to other degrees of freedom, would produce Cherenkov radiation, which would slow down the particles, thus preventing them from reaching the instability point. 

The inhomogeneous, similarity-gauged non-Hermitian models also possess, again at sufficiently high momenta, negative group velocities, and anomalous dispersion relations. Moreover, at a certain momentum, the group velocity of one of the modes vanishes. All such exotic effects (apparent superluminality, negative group velocity, and anomalous dispersion), found for the two-flavour non-Hermitian field theories analysed in this work, have their classical counterparts in allowed propagation through an absorbing medium~\cite{Garrett1970}, which calls for future investigation of the parallels between these physical systems.

Despite the non-Hermitian nature of the effect brought by inhomogeneities and notwithstanding the exotic features of the system at high energies, the propagating modes reside in the $\PT$-unbroken domain at low energies. In particular, the physical excitations at low momenta have a real-valued spectrum, implying that the non-Hermitian modifications do not manifest in the experimentally accessible low-energy domain. We leave detailed phenomenological studies for future work.

%%%%%%%%%%%%%%%%%%%%

\acknowledgments

The authors thank Esra Sablevice for useful discussions. This work was supported by a Royal Society International Exchange [Grant No.~IES\textbackslash R3\textbackslash203069]. The work of P.M.~was supported by the Science and Technology Facilities Council (STFC) [Grant No.~ST/X00077X/1] and a United Kingdom Research and Innovation (UKRI) Future Leaders Fellowship [Grant Nos.~MR/V021974/1 and~MR/V021974/2]; and a Nottingham Research Fellowship from the University of Nottingham. For the purpose of open access, the authors have applied a Creative Commons Attribution (CC BY) licence to any Author Accepted Manuscript version arising.

%%%%%%%%%%%%%%%%%%%%

\section*{Data Access Statement}

No data were created or analysed in this study.

%%%%%%%%%%%%%%%%%%%%

\appendix

%%%%%%%%%%%%%%%%%%%%

\section{Momentum-dependent inner products}
\label{sec:innerprod}

In this Appendix, we describe the momentum-dependence of the positive-definite $\mathcal{APT}$ inner product induced by a non-vanishing similarity gauge field $C^{\mu}$. For simplicity, we will consider constant $C^{\mu}$.

The momentum-dependence is most straightforwardly studied at the level of the $2\times 2$ flavour space. Therein, the effective squared Hamiltonian for both the two-flavour scalar and fermion models takes the general form:
\begin{equation}
    H^2(\bs{p})=\begin{pmatrix} \bs{p}^2+m_1^2-C^2 & m_3^2-i m_C^2(\bs{p}) \\ -m_3^2-i m_C^{2}(\bs{p}) & \bs{p}^2+m_2^2-C^2
    \end{pmatrix},
\end{equation}
where $m_3^2$ and $m_C^2(\bs{p})$ are, in general, functions of the similarity gauge field $C^{\mu}$. The dependence of $m_C^2(\bs{p})$ on the three-momentum $\bs{p}$ is induced by the first-order spatial derivatives in the equations of motion.

We proceed as in the main body by separately treating the timelike $C^\mu=(C_0,\bs{0})$ and spacelike $C^\mu=(0,\bs{C})$ cases and make the following correspondences for the two scalar (upper element of the braces) and two fermion models (lower element of the braces):
\begin{align*}
    \text{timelike case:}\quad &C=C_0\;,\\ &m_1^2=\begin{Bmatrix} m_1^2\\ M_0^2+m_0^2\end{Bmatrix}\;,\\ 
    &m_2^2=\begin{Bmatrix} m_2^2 \\ M_0^2-m_0^2\end{Bmatrix}\;,\\
    &m_3^2=\begin{Bmatrix}[m_5^4+2C_0^2(m_1^2+m_2^2)]^{1/2}\\ [m_0^4+M_0^2(C_0^2-4m_0^2)]^{1/2}\end{Bmatrix}\;,\\ 
    &m_C^2(\bs{p})= 2\, C_0|\bs{p}|\;,\\
    \text{spacelike case:}\quad &C=|\bs{C}|\;,\\
    &m_1^2=\begin{Bmatrix} m_1^2\\ M_0^2+m_0^2\end{Bmatrix}\;,\\
    &m_2^2=\begin{Bmatrix} m_2^2 \\ M_0^2-m_0^2\end{Bmatrix}\;,\\
    &m_3^2=\begin{Bmatrix}m_5^2\\ [m_0^4+4m_0^2(|\bs{C}|^2-M_0^2)]^{1/2}\end{Bmatrix}\;,\\
    &m_C^2(\bs{p})=2\,\bs{C}\cdot \bs{p}\;.
\end{align*}

The parity matrix $P={\rm diag}(1,-1)$ is unchanged from the case of the vanishing similarity gauge field. However, the non-Hermitian parameter is given by
\begin{equation}
    \zeta\equiv \zeta_C(\bs{p})=\frac{2[m_3^2-i m_C^2(\bs{p})]}{m_1^2-m_2^2}\in \mathbb{C}\;,
\end{equation}
and the eigenvectors are now complex, taking the form
\begin{subequations}
    \begin{align}
\mathbf{e}_+=\mathcal{N}\begin{pmatrix}\zeta \\ -1+\sqrt{1-|\zeta|^2} \end{pmatrix},\\
\mathbf{e}_-=\mathcal{N}\begin{pmatrix}-1+\sqrt{1-|\zeta|^2} \\ \zeta^* \end{pmatrix},
\end{align}
\end{subequations}
with normalisation
\begin{equation}
    \mathcal{N}=\left[2\left(|\zeta|^2-1+\sqrt{1-|\zeta|^2}\right)\right]^{-1/2}.
\end{equation}
Notice that the normalisation $\mathcal{N}$ remains real-valued in the $\mathcal{PT}$-symmetric regime. We also see that the non-Hermitian parameter is momentum-dependent for a non-vanishing similarity gauge field, as we might expect from the presence of the momentum-dependent exceptional points identified in the main text.

The matrix $A$ appearing in the $\mathcal{APT}$ inner product becomes
\begin{equation}
    A=\frac{1}{\sqrt{1-|\zeta|^2}}\begin{pmatrix} 1 & \zeta \\ -\zeta^* & -1\end{pmatrix}\;,
\end{equation}
and the product
\begin{equation}
    P\cdot A=\frac{1}{\sqrt{1-|\zeta|^2}}\begin{pmatrix} 1 & \zeta \\ \zeta^* & 1\end{pmatrix}
\end{equation}
gives a Hermitian matrix, as it should form the relevant positive-definite inner product for this non-Hermitian theory.

It can readily be confirmed that the eigenvectors are orthonormal with respect to this inner product:
\begin{subequations}
    \begin{align}
        \mathbf{e}_{\pm}^*\cdot P\cdot A\cdot \mathbf{e}_{\pm}=1\;,\\
        \mathbf{e}_{\pm}^*\cdot P\cdot A\cdot \mathbf{e}_{\mp}=0\;.
    \end{align}
\end{subequations}
Moreover, we can also confirm that
\begin{equation}
        \mathbf{e}_{\pm}^*\cdot P\cdot A\cdot H^2(\bs{p})\cdot\mathbf{e}_{\pm}=\omega_{\pm,\bs{p}}^2\;.
\end{equation}
Naively, we might expect the momentum-dependence of the matrix $A$ to give rise to subtleties with respect to the calculation of the group velocity. However, we find that this is not the case, since
\begin{align}
    \bs{v}_{g,\pm,\bs{p}}&=\frac{1}{2\omega_{\pm,\bs{p}}}\,\nabla_{\bs{p}}\omega_{\pm,\bs{p}}^2\nonumber\\&=\frac{1}{2\omega_{\pm,\bs{p}}}\,\mathbf{e}_{\pm}^*\cdot P\cdot A\cdot [\nabla_{\bs{p}}H^2(\bs{p})]\cdot\mathbf{e}_{\pm}\;,
\end{align}
and we are justified in focusing our attention on the dispersion relation as derived in the main body.

%%%%%%%%%%%%%%%%%%%%

\section{Non-Hermiticity and the electromagnetic sector}
\label{sec:emcharges}

The group of similarity transformations, either global or local ones, commutes with the group of electromagnetic gauge transformations given by
\beqn
\begin{array}{l}
{\mathrm{Electromagnetic}} \\ 
U_{\mathrm{e.m.}}(1)\ {\mathrm{gauge}} \\
{\mathrm{transformation}}
\end{array}
\!:\quad 
\left\{
\begin{array}{lcl}
\Psi(x) & \to & e^{i e \alpha(x)} \Psi(x) \\ 
\bar \Psi(x) & \to & e^{- i e \alpha(x)} \bar \Psi(x) \\
A_\mu(x) & \to & A_\mu(x) + \partial_\mu \alpha(x)
\end{array}
\right.\!\!\!\!,
\qquad
\label{eq_electromagnetic:gauge}
\eeqn
provided that the electric charges of both components $\psi_1$, and $\psi_2$ of the fermion doublet $\Psi = (\psi_1,\psi_2)^T$ possess the same electric charges, $e_1 = e_2 \equiv e$.

The covariant derivative~\eq{eq_D_mu} in the non-Hermitian model~\eq{eq_L:gauged} can be extended to include also the electromagnetic gauge field $A_\mu$ as follows:
\beqn
D_\mu = \bbbone \, \partial_\mu + \sigma_1 C_\mu \to \cD_\mu = \bbbone \, (\partial_\mu - i e A_\mu) + \sigma_1 C_\mu.
\qquad
\label{eq_D:shift:1}
\eeqn

The electric current, corresponding to the variation of the action with respect to the gauge field $A_\mu$, is given by the standard expression:
\beqn
j_{\mathrm{e.m.}}^\mu = e {\bar\Psi} \gamma^\mu \Psi \equiv e \sum_{a=1}^2 {\bar \psi}_a \gamma^\mu \psi_a, \qquad [e_1 = e_2 = e].
\qquad
\label{eq_EM}
\eeqn
According to the classical equation of motion~\eq{eq_EoM}, the electric current~\eq{eq_EM} is a classically conserved quantity both in the original~\eq{eq_L} and the similarity-gauged~\eq{eq_L:gauged} versions of the non-Hermitian model, i.e.,
\beqn
\partial_\mu j_{\mathrm{e.m.}}^\mu = 0\;.
\eeqn

If the charges of the two components differ, i.e., $e_1 \neq e_2$, then either the global or local $SU(1,1)$ group of similarity transformations~\eq{eq_similarity:matrix} becomes broken explicitly by the Maxwell $U(1)$ group of the electromagnetic gauge transformations:
\beqn
U = e^{i e {\hat I}_{\mathrm{e.m.}} \alpha}\,, \qquad
{\hat I}_{\mathrm{e.m.}} = \frac{e_1 + e_2}{2e}\bbbone + \frac{e_1-e_2}{2e}\sigma_3. \qquad
\eeqn
If $e_1 \neq e_2$ then the generator of the electromagnetic group ${\hat I}_{\mathrm{e.m.}}$ does not commute with the $\sigma_1$ generator of the Cartan subgroup of the $SU(1,1)$ similarity group, $[{\hat I}_{\mathrm{e.m.}},\sigma_1] \neq 0$, and the similarity group is explicitly broken by the difference in the electromagnetic charges.  The covariant derivative~\eq{eq_D:shift:1} then reads as follows:
\beqn
\cD_\mu = \bbbone \, \partial_\mu - i e {\hat I}_{\mathrm{e.m.}} A_\mu + \sigma_1 C_\mu\,.
\qquad
\label{eq_D:2}
\eeqn

In order to maintain the gauge invariance of this theory, the  mass $m_5$ should be promoted to a field (condensate), $\phi_5$, in the off-diagonal mass term in Eq.~\eq{eq_L:gauged:explicit}:
\beqn 
m_5 {\bar \psi}_1 \psi_2 \to {\bar \psi}_1 \phi_5 \psi_2, 
\qquad
m_5 {\bar \psi}_2 \psi_1 \to {\bar \psi}_2 \phi_5^* \psi_1.
\eeqn
The field $\phi_5$ should transform under $U(1)$ electromagnetic group as follows:
\beqn
U_{\mathrm{e.m.}}(1), \qquad \phi_5 \to e^{- i (e_1 - e_2) \alpha} \phi_5\,,
\eeqn
where the gauge transformation parameter $\alpha$ is the same as in Eq.~\eq{eq_electromagnetic:gauge}.
The diagonal masses $m_1$ and $m_2$ remain gauge invariant. However, in this case, the electric current, which can be read off from the expression for the covariant derivative~\eq{eq_D:2},
\beqn
j_{\mathrm{e.m.}}^\mu = e {\bar\Psi} \gamma^\mu {\hat I}_{\mathrm{e.m.}} \Psi, \qquad [e_1 \neq e_2],
\qquad
\label{eq_EM:2}
\eeqn
becomes a non-conserved quantity:
\beqn
\partial_\mu j_{\mathrm{e.m.}}^\mu & = & (e_2 - e_1) \left[ {\bar \psi}_1 (m_5 {+} {\slashed C}) \psi_2 
{+} {\bar \psi}_2 (m_5 {-} {\slashed C}) \psi_1 \right] \qquad \\
 & \equiv & (e_2 - e_1) \, {\bar \Psi} (m_5 - i \sigma^2 {\slashed C}) \Psi \,.
\label{eq_non-conservation}
\eeqn
Notice that the source of the non-conservation, given on the right-hand side of this expression, is a Hermitian quantity. The non-conservation of the electric charge~\eq{eq_non-conservation} makes phenomenologically questionable the fermionic non-Hermitian models with unequal charges ($e_1 \neq e_2$), thereby forcing us to consider the fermionic doublets with equal charges ($e_1 = e_2$).

\end{document}